\documentclass[aps,prl,twocolumn,nopacs,superscriptaddress,longbibliography]{revtex4-1}

\usepackage[breaklinks=true,colorlinks,citecolor=blue,linkcolor=blue,urlcolor=blue]{hyperref}
\usepackage{epsfig,mathrsfs,color,latexsym,subfigure,marginnote,graphicx,verbatim,relsize,mathrsfs,color,array,amsmath,amsfonts,amssymb,graphicx}
\begin{document}

\title{ High-order Van Hove singularities in cuprates and related high-$T_c$ superconductors }

\author{Robert S. Markiewicz}
\email{r.markiewicz@northeastern.edu}
\affiliation{Department of Physics, Northeastern University, Boston, Massachusetts 02115, USA}

\author{Bahadur Singh}
\email{bahadur.singh@tifr.res.in}
\affiliation{Department of Condensed Matter Physics and Materials Science, Tata Institute of Fundamental Research, Colaba, Mumbai 400005, India}

\author{Christopher Lane}
\affiliation{Theoretical Division, Los Alamos National Laboratory, Los Alamos, New Mexico 87545, USA}
\affiliation{Center for Integrated Nanotechnologies, Los Alamos National Laboratory, Los Alamos, New Mexico 87545, USA}

\author{Arun Bansil}
\affiliation{Department of Physics, Northeastern University, Boston, Massachusetts 02115, USA}

\begin{abstract}

Two-dimensional (2D) Van Hove singularities (VHSs) associated with the saddle points or extrema of the energy dispersion usually show logarithmic divergences in the density of states (DOS). However, recent studies find that the VHSs originating from higher-order saddle-points have faster-than-logarithmic divergences, which can amplify electron correlation effects and create exotic states such as supermetals in 2D materials. Here we report the existence of `high-order' VHSs in the cuprates and related high-$T_c$ superconductors and show that the anomalous divergences in their spectra are driven by the electronic dimensionality of the system being lower than the dimensionality of the lattice. The order of VHS is found to correlate with the superconducting $T_c$ such that materials with higher order VHSs display higher $T_c’s$. We further show that the presence of the normal and higher-order VHSs in the electronic spectrum can provide a straightforward marker for identifying the propensity of a material toward correlated phases such as excitonic insulators or supermetals. Our study opens up a new materials playground for exploring the interplay between high-order VHSs, superconducting transition temperatures and electron correlation effects in the cuprates and related high-$T_c$ superconductors. 
\end{abstract} 

\maketitle

{\it Introduction.$-$} The diversity of electronic properties in materials often originates from their one-particle electronic density of states (DOS). The large DOS at the Fermi level generally implies that many electrons contribute to the low-energy phenomena so that many-body interactions are enhanced. In particular, extrema and saddle-points in band dispersion induce Van Hove singularities (VHSs) in the DOS \cite{VHS}.  Such VHSs in two-dimensions (2D) normally lead to a logarithmically diverging DOS and have been a focus of interest for many years. Recent studies show that the 2D VHSs can be anomalously strong with a power-law divergence \cite{Liang, Liang3, MBMB, Marsig}. Such VHSs can enhance many-body interaction and drive more exotic correlated phenomena such as supermetals \cite{Liang2}. The high-order VHSs have been reported in materials with flat bands such as Moire heterostructures, slow-graphene, magic-angle twisted bilayer graphene, Sr$_3$Ru$_2$O$_7$, among other systems \cite{Liang,Tgraphene,SRO327}. Undoubtedly the reduced bandwidth in these systems leads to dominant Coulomb interactions which can drive instabilities toward various correlated states. The added electronic feature of reduced bandwidth is an enhanced DOS which assures stronger electron correlations. Although it is clear that high-order VHSs hold great promise for driving various correlated states, merely classifying their DOS anomalies is not sufficient for understanding the complex effects associated with these VHSs. 

The role of normal VHSs in the cuprates and related high-$T_c$ superconductors has been debated for decades. One of the earliest theories of cuprate superconductivity is that it is driven by a large DOS associated with the VHSs \cite{HirSc}. However, the actual relationships between the cuprate superconductivity, doping, and VHSs are more complex and material dependent. The doping at both optimal superconductivity $x_{\mathrm{SCmax}}$ and pseudogap collapse $x_{\mathrm{pg}}$ satisfy \{$x_{\mathrm{SCmax}},x_{\mathrm{pg}}\}\le x_{\mathrm{VHS}}$, where $x_{\mathrm{VHS}}$ is the doping at which the VHS crosses the Fermi level. For the lanthanum-based cuprates, these inequalities become equalities, leading to high-order VHSs at the Fermi level near optimal doping. Importantly, such a high-order VHS may have been observed recently near $x_{pg}$ in the cuprates \cite{Michon}, consistent with an earlier prediction \cite{RM70}. For other cuprates, there can be a larger gap between $x_{VHS}$ and the superconducting or pseudogap doping \cite{Lizaire2020}. Regardless, an analysis of the order of VHS evolution and the resulting anomalies in the cuprates may bring new insights for understanding and identifying new correlated states.

In this work, we explore the existence of high-order VHSs in cuprate high-$T_c$ superconductors and illustrate how these VHSs can drive complex effects and various competing orders.  We numerically confirm the slope quantization by considering reference cuprate energy dispersions.  We find a VHS dichotomy where the singularities exist not only in the DOS or $Q=(0,0)$ susceptibility but also at a finite $Q\sim(\pi,\pi)$ momentum. These singularities compete with each other and show independent evolutions with tuning the dispersions or with doping away from the VHSs. We show how tuning these dispersions can generate flat bands with high-order VHSs that create frustration rather than instability. We also show the existence of high-order VHSs in bosonic bands and discuss that if the susceptibility is considered as a dispersion of electron-hole pair Bosons, the resulting high-order Bosonic VHSs could resemble secondary electronic VHSs, which would play an important role in a straightforward identification of excitonic phases in materials.

\begin{figure}[h!]
\centering
\includegraphics[width=0.99\columnwidth]{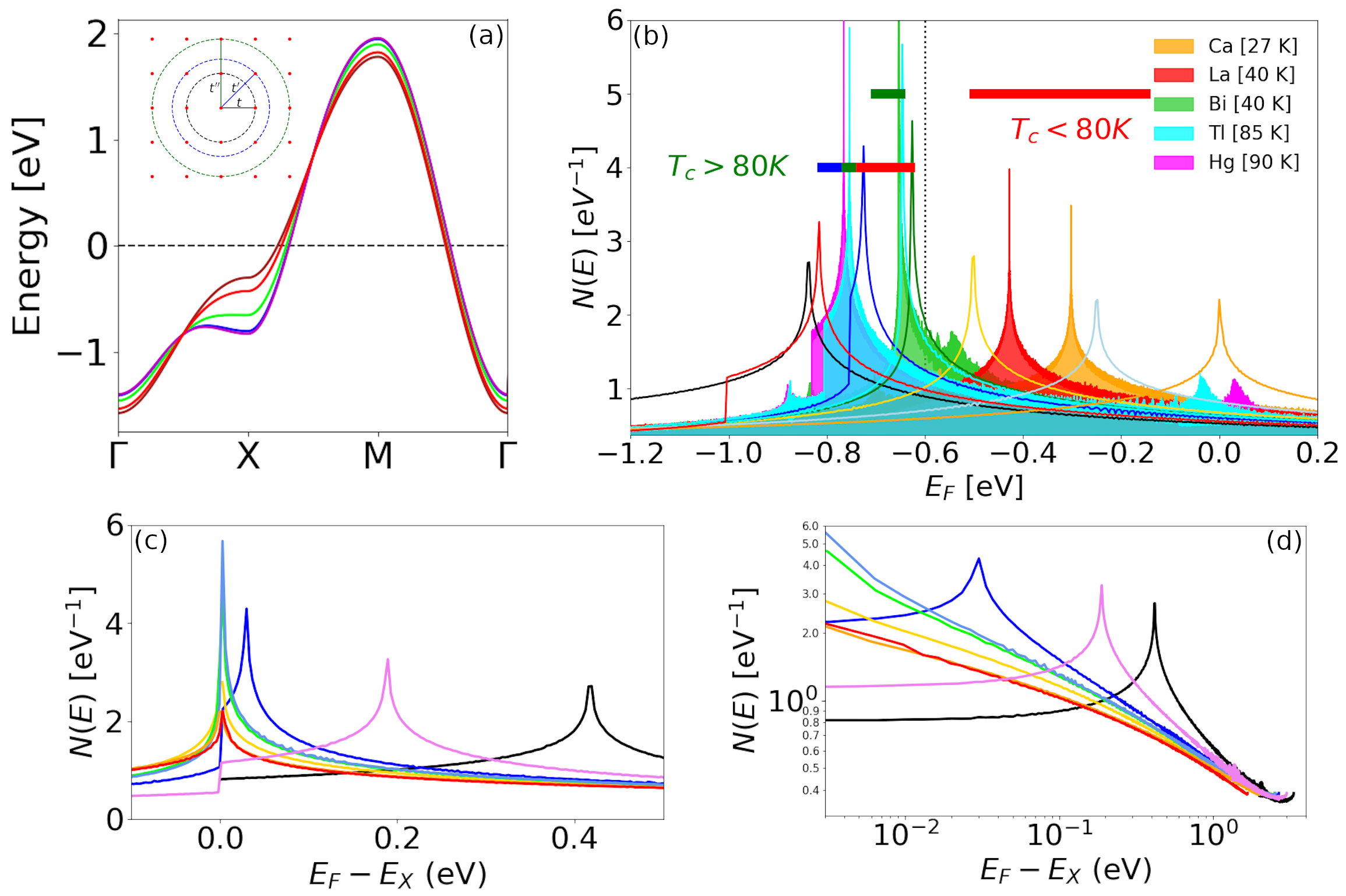}
\caption{{\bf Reference family of the cuprates ($t''=-t'/2$).} (a) Dispersions for the five cuprates of frame (b).  Inset: Definition of the hopping parameters $t$, $t'$, and $t''$.  (b) DOS $N(E)$ for several values of $t'$.   As VHS moves from right to left, the white-background curves correspond to $t'/t$ = 0 (red curve), -0.1 (orange), -0.2 (yellow-green), -0.25 (green), -0.258 (light blue), -0.3 (blue), -0.4 (violet),  and -0.5 (black), while the colored-background curves correspond to the monolayer cuprates as indicated in the legends with $t'$ values from Ref.~\onlinecite{PavOK}.  Horizontal bars indicate range of VHS peak positions for 10 bilayer or trilayer cuprates\cite{PavOK}, sorted by optimal $T_c$s, with red ($70K\ge T_c\ge 50K$), green ($100K\ge T_c\ge 90K$), and blue ($135K\ge T_c\ge 125K$) colors. The antibonding bands are indicated by thick bars and the bonding bands by thin bars. A clear correlation of high-order VHSs with higher superconducting $T_c$s is seen.  (c,d) White-background data from frame (b) replotted as $E_F-E_X$, on linear (c) or logarithmic (d) scales.We consider an average DFT value of $t=-0.5$ eV for all the calculations.} 
\label{fig:1vhs} 
\end{figure}

{\it High-order VHSs$-$} We begin by recalling that the energy dispersion in the cuprates can be described by a $t-t'-t''$ model \cite{MBMB}

\begin {equation}
E=-2t(c_x+c_y)-4t'c_xc_y-2t''(c_{2x}+c_{2y}),
\end{equation}
where $c_{nr} = cos(nk_ra)$, $a$ is the lattice constant, $r = \{x, y\}$, and the hopping parameters are defined in the inset to Fig.~\ref{fig:1vhs}(a).  In this model, $t$ sets the energy scale and thus, evolution of energy dispersions depends on the two parameters, $t'/t$ and $t''/t$ which constitute the material dependence. We examine this energy dispersion considering $t'/t$ and $t''/t$ values relevant to cuprates and delineate high-order VHSs. It should be noted that nearly localized $d$- and $f$ electrons may be sensitive to just a few hopping parameters so that similar calculations should determine the characteristic properties of high-order VHSs in many correlated materials beyond cuprates.

Figure \ref{fig:1vhs}(b) presents the evolution of VHS lineshapes with $t'$ for the special value $t''=-t'/2$ that best describes most families of cuprates \cite{PavOK}.  The corresponding VHS peak is considered as a marker of $t'$ which locates various cuprates along the $x$-axis similar to Ref.~\onlinecite{PavOK} where the $r \sim t'/t$ parameter correlates materials with their superconducting $T_c$. Such a correlation of $t'$ with the VHS allows us to compare the strength of the VHS divergence with $T_c$ for several families of cuprates. The vertical dotted line in Fig.~\ref{fig:1vhs}(a) separates the VHS peaks into a logarithmic shape at small values of $t'$ from the stronger divergence ($\sim$power law) at large values of $t'$. Strikingly, this dotted line also separates cuprates with $T_c \le 80K$ from those with $T_c> 80K$, with one exception.  The Bi-cuprates are the only cuprate family in which $T_c$ changes significantly with number of CuO$_2$-layers per unit cell, even though all have similar $t'$-values.  Hence Bi2201, with maximum $T_c$ of 40K, ends up just on the other side of the crossover line.  Note further that for bilayer cuprates, the correlation holds for the antibonding band that is closer to the Fermi level, whereas the bonding bands all have $t'$s that correspond to super VHSs.  These results clearly suggest that high-order VHSs play a significant role in the superconductivity of cuprates.
 
To further understand the evolution of VHSs, it is convenient to measure energy $E$ from the energy of the $X =(\pi,0)$ point $i. e.$ $E_X=4 (t'-t'')$. As seen in Fig.~\ref{fig:1vhs}(c), there is always a VHS at $E_X$, which evolves from logarithmic (saddle-point) at small $t'$ to the step at larger $t'$.  The associated single Fermi pocket region changes into a region of Fermi surface with three pockets. The step is the point at which two pockets first appear for a given $t'$.   The crossover occurs at a critical value $t'_c$ where a pocket forms with strongest VHS divergence (for that $t''/t'$). It has a step on the low-energy side and a power-law divergence on the high energy side.  This evolution is further illustrated by replotting the data for $E>E_X$ on logrithmic scales in Fig.~\ref{fig:1vhs}(d).  There are two types of behavior, separated by the turquoise line.  For small $|t'|$ (red to turquoise curve, the divergent peak stays at $E_X$, evolving from logarithmic to power law.  The turquoise curve has the largest, pure power law divergence.  For larger $|t'|$, all curves (black to turquoise) start off with the same power law growth at large $\delta E=E_f-E_X$, but as $\delta E$ decreases, the curves gradually split off on realizing a power-law to  logarithmic crossover at an energy away from $E_X$. Finally, we note from Fig.~\ref{fig:1vhs}(a) that the $t'$ of strongest VHS can be approximately determined from the dispersion, as corresponding to the flattest band near $(\pi,0)$.  Such strong VHS corresponds to Andersen's extended VHS \cite{Andex}.

\begin{figure}[h!]
\centering
\includegraphics[width=0.99\columnwidth]{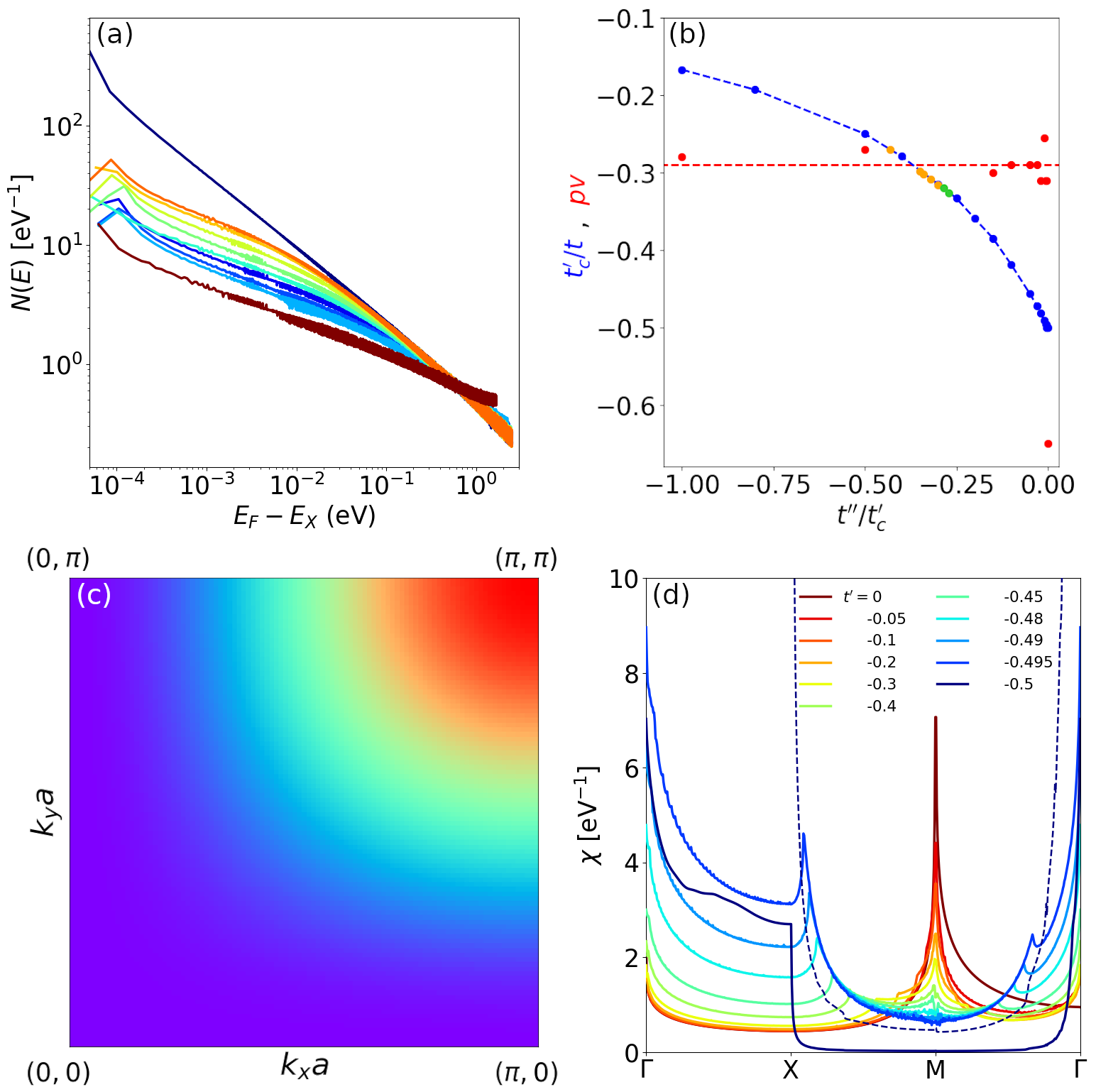}
\caption{(a) {\bf Power-law divergence of $N(E)$ at $t'_c$} for various $t''/t'$ =(from top to bottom)  0.0 , -0.05, -0.10, -0,15, -0.20, -0.25, -0.258 -0.30, -0.35, -0.40, -0.45, and -0.50.  (b) $t'_c$  (blue line) and power-law exponent $p_v$ (red filled circles) vs $t''/t'$.  (c) Dispersion for $t''=0$, $t'=-t/2$.  (d) Susceptibility along high-symmetry directions for $t-t'$ reference family.  For $t'=-0.5t$, the blue dashed curve gives the susceptibility while the solid blue curve is the susceptibility divided by 20. 
}
\label{fig:2Suscept}
\end{figure}

The above behavior is universal in two ways. Firstly, as $\delta E$ is reduced and $t'$ moves closer to $4t'_c$, the logarithmic (ln) plot becomes a scaled version of Fig.~\ref{fig:1vhs}(c). Secondly, the same pattern is repeated for almost all ratios $t''/t$ that we have studied. This indicates that for each reference family (specified by $t''/t'$), there is a specific $t'_c$ (Figure~\ref{fig:1vhs}(b)) at which a high-order VHS exists with power-law divergence. Figure~\ref{fig:2Suscept} illustrates this divergence for a series of ratios $t''/t'$ in the range 0 to -0.5. While the black curve ($t''=0$) has a single power-law divergence with slope $p_V=-0.65$, all other curves have two regions of different power-law divergence. Remarkably, all curves converge to the $t''=0$ curve at higher values of $\delta E$ whereas they shift to a weaker divergence at low $\delta E$ with approximately the same power-law $p_V=-0.29$ (Fig.~\ref{fig:2Suscept}(b)).  The calculation of $p_V$ is further discussed in Supplementary Material (SM).

We now show that high-order VHSs exist if the electronic dimensions are smaller than lattice dimensions. During the early days of many-body perturbation theory (MBPT), it was postulated that the effective electron dimensionality could be smaller than the crystal lattice dimension {\it i.e.} if a Fermi surface has flat parallel sections there would be good nesting, leading to quasi-1D behavior.  For example, the early high-T$_c$ superconductors such as the A15 compounds were assumed to be composed of three orthogonal interpenetrating chains of electrons \cite{Labbe}. Figure~\ref{fig:2Suscept} provides clear evidence that the most singular high-order VHS for each choice of $t''$ is dominated by quasi-one-dimensionality of the electrons, and that frustration causes the divergence to weaken.  To understand the origin of the one-dimensionality, it is convenient to look at the dispersion of the state with the strongest instability, corresponding to $t''=0$ and $t'=-t/2$ in Fig.~\ref{fig:2Suscept}(c).  For these parameters, $E=-2t[c_y+c_x(1-c_y)]=-2t$ for $k_y=0$ which is independent of $k_x$.  This dispersion is thus flat along the $y$-axis as well as along the $x$-axis due to symmetry.  Despite this, the susceptibility is not uniform along the $y$-axis due to the crossing of the $x$-axis susceptibility at $\Gamma$ (see Fig.~\ref{fig:2Suscept}(d)). More specifically, recall that a 2D saddle-point VHS corresponds to a point in k-space where the local dispersion has the form $ak_x^2-bk_y^2$, leading to a logarithmic peak in the DOS while  a high-order VHS arises when, for example, $b\rightarrow 0$.  In contrast, a 1D VHS arises when $b=0$ over an extended line segment.  In present case, when $k_y$ is small the dispersion has the form $E = b(k_x)k_y^2$ along the whole $k_x$-axis, leading to a 1D VHS.  Moreover, $b\sim 1-cos(k_xa)\rightarrow 0$ as $k_x\rightarrow 0$, leading to a 1D high-order VHS with $p_V=0.65>0.5$, the conventional 1D result.

The susceptibility is largest at $\Gamma=(0,0)$, corresponding to the DOS.   The lineshape of $N(E)$ is extremely asymetric since the VHS falls at the bottom of the band. The DOS has a step from zero to infinity on one side, and the power-law fall-off on the other side (Fig.~\ref{fig:2Suscept}(a)).  For smaller $|t'|$, the susceptibility decreases rapidly in  Fig.~\ref{fig:2Suscept}(d), and the DOS at the VHS reverts to the conventional logarithmic form expected for 2D electrons.  We note in passing that since the dispersion is flat along the $x$ and $y$-axes, it cannot be represented by any function of the form $f_1(k_x)-f_2(k_y)$. A finite $t''>0$ modulates the dispersion along the axes, greatly weakening the divergence. The strongest residual divergence arises at the point when $(\pi,0)$ pockets first form.   The DOS retains a quasi-1D lineshape with a power-law divergence with a weaker power $p_V$ on one side and a step down on the other side.

{\it VHS Dichotomy and Secondary VHSs.$-$} The instabilities associated with the VHSs form a Lie group which is SO(8) for cuprates.\cite{Mikefest} For our purpose, the most important subgroup is whether the instability involves intra-VHS coupling which corresponds to $q=0$ {\it i.e.}, a peak in the DOS or with inter-VHS coupling, producing a peak in the $Q=(\pi,\pi)$ susceptibility. These two instability modes compete  such that in the original Hubbard model the $(\pi,\pi)$ instability dominates at half-filling leading to a $ln^2$ instability. Thus, focusing solely on the DOS would miss the strong antiferromagnetism of cuprates. The generality of this $ln^2$ effect has been questioned since the Hubbard model requires extreme fine-tuning with all hopping parameters set to zero except nearest-neighbor $t$. However, in SM  II we display a large family of dispersions with $ln^2$ susceptibility divergence.

Remarkably, the $(\pi,\pi)$ VHS is completely insensitive to the Fermi surface nesting that produces structure in the DOS, only gradually crossing over from $ln^2$ to $ln$ as the dispersion is tuned away from the Hubbard limit by either doping or tuning hopping parameters.  Hence in general at some hopping $t'_{cross}$ the DOS instability will become dominant.  This is accompanied by an $x_{cross}$ where the dominant near-$(\pi,\pi)$ instability crosses over to a near-$\Gamma$ instability (see SM II).  We believe that in most cuprates this crossover at $x_{cross}$ plays a larger role than the doping $x_{VHS}$.  For instance, this is where AFM order crosses over to stripe or charge-density wave phases. We further suggest that optimal superconductivity falls close to $x_{cross}$.  That is because an electron-electron driven instability such as superconductivity is at a disadvantage compared to an electron-hole instability such as AFM or CDW.  However, when two e-h instabilities are competing, superconductivity can tilt the balance, acting as a symbiotic parasite.  Also, fluctuations will be large near $x_{cross}$ and can further enhance $T_c$.

\begin{figure}[h!]
\centering
\includegraphics[width=0.99\columnwidth]{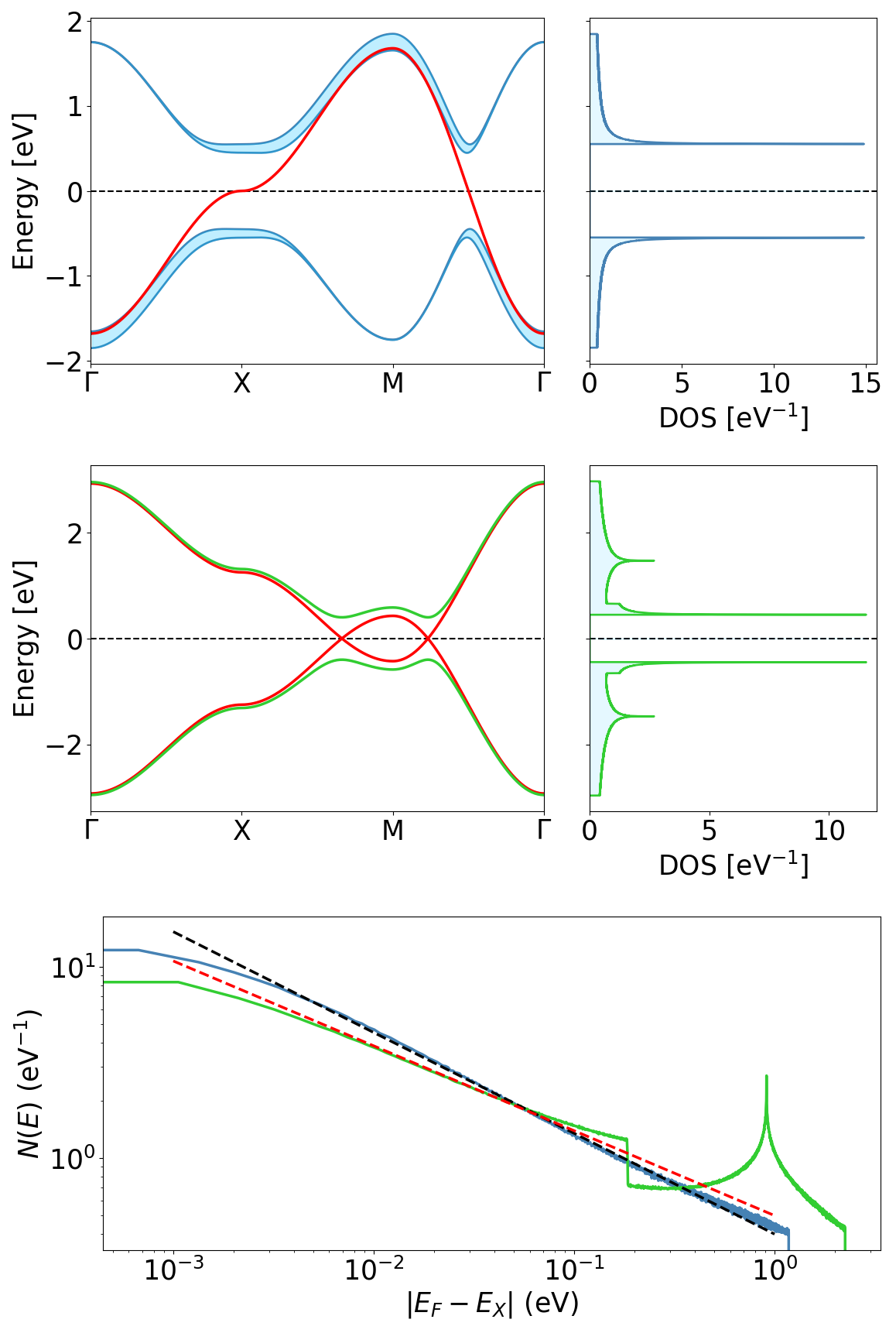}
\caption{{\bf Emergence of secondary VHSs} (a) $(\pi,\pi)$ Antiferromagnetic Hubbard model $(t'=t''=0)$ at half-filling with bare dispersion (red line) and gapped dispersion with mean-field gap parameter $\Delta= 0.5$~eV (blue curves). The width of the blue curve indicates spectral intensity. (b) The associated DOS. (c) Excitonic insulator model with bare dispersion (red line) and gapped dispersion with mean-field gap parameter $\Delta= 0.4$~eV (green curves) and (d) the resulting DOS. (e) Panels (b) and (d) are replotted on a {\it ln-ln} plot to show the high-order VHSs. The solid and dashed black lines provide the reference slopes of -0.54 and -1/2, respectively.}
\label{fig:3b}
\end{figure}

The above analyses do not exhaust the possibilities for high-order VHSs. They are so singular that they could drive an electronic phase transition at high temperatures to open gap in the electronic spectrum. However, the gap does not destroy the VHS, but replaces it by a pair of saddle-point VHSs, one for each band created by the gap opening. Hence electronic phase transitions are likely to be powerful sources of exotic secondary VHSs, with properties that could be quite different from the primary VHSs discussed above. Here we provide the two examples of such secondary VHSs.  
Firstly, let us consider the DOS in the mean-field antiferromagnetic (AFM) phase in a pure Hubbard model ($t'=t''=0$), where the secondary VHS displays strong frustration.  Figures~\ref{fig:3b}(a) and (b) show that DOS has a strong power-law divergence although it is not associated with 1D nesting.  Instead, the associated dispersion is exactly balanced at a crossover from having a dispersion minimum (in the upper band) at $(\pi,0)$ to having a local maximum, so the whole dispersion is drumhead-flat, leading to the anomalously large DOS.  Notably, when $t'$ is non-zero this evolves into a Mexican hat dispersion, with local maximum at $(\pi,0)$, and resulting strong frustration \cite{RSMstripe}.  When the AFM gap closes with increased doping, the high-order VHS of the lower magnetic band merges into the step at the bottom of the upper magnetic band, to form the high-order VHS of the nonmagnetic band (see Ref.~\onlinecite{RSMstripe} and Fig.~17 of Ref.~\onlinecite{RM70}).  Such a feature was recently seen experimentally\cite{Michon}, but the VHS interpretation was discarded because the feature was too intense to be a conventional logarithmic VHS \cite{Michon,Horio}.

Secondly, we investigate the excitonic insulator model where electron and hole pockets bind together to form avoided crossings at the Fermi level. We consider electron and hole pockets of the same size and shape to overlap at the same k point.  Notably, the electron and hole pockets can have different geometries or be shifted by a fixed $q$. However, for simplicity we consider these pockets of the same area and shape. Adding a hybridization term to such a model leads to an avoided crossing with a `Mexican-hat' dispersion (Fig.~\ref{fig:3b}(c)). From Fig.~\ref{fig:3b}(d) it is seen that this dispersion leads to a highly characteristic DOS in either hybridized band. The logarithmic VHSs of the original bands (at $\pm 1.5$eV) remain unchanged by hybridization whereas the band edge VHSs split into two VHSs- a step and a power-law VHS asone move towards the band edge. The power-law saddle points are associated with an unusual Higgs-like one-dimensionality in the bands where they are flat along the brim of the Mexican hat but form an extremum along in the radial $k$-direction away from $(\pi,\pi)$. Similar avoided crossings are found in TiSe$_2$\cite{singh2017} and often seen in the topological insulators with a band inversion \cite{Bansil2016}.    

\begin{figure}[h!]
\centering
\includegraphics[width=0.99\columnwidth]{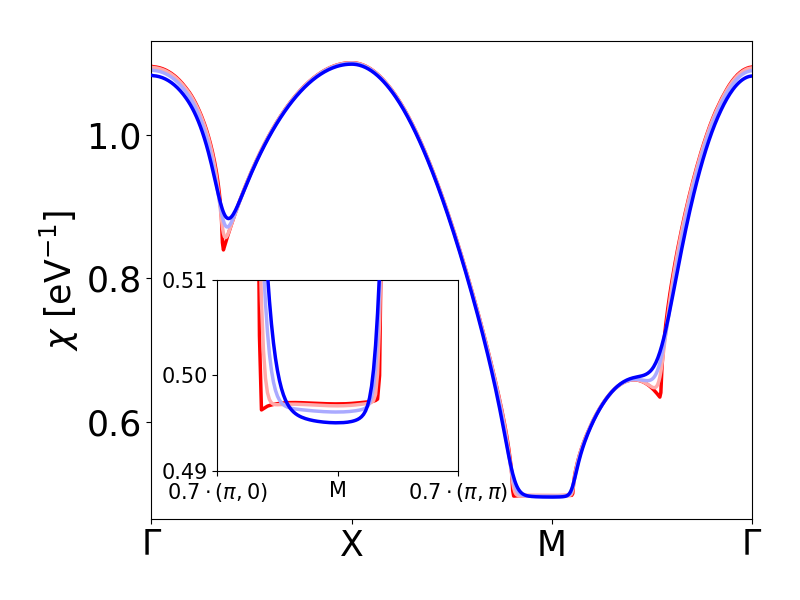}
\caption{{\bf Cuprate bosonic dispersion $\omega_q$.}    Bare dispersion (red line) and gapped dispersion with mean-field gap parameter  $\Delta= 0.5$~eV (blue curves).  Inset: Blowup near band minimum.}
\label{fig:3c}
\end{figure}

We emphasize that these VHSs can be dubbed as Overhauser VHSs \cite{Over} since their model of charge density waves involves singular interactions on a 3D spherical Fermi surfaces. The circular VHSs would therefore provide a realistic 2D version of their effect. This is a particular example where a flat band leads to strong frustration which can greatly lower the transition temperature. We find in Fig.~\ref{fig:3c}(e) that the secondary VHSs have distinct power laws ($p_V = 0.5, 0.54$) from the primary VHSs ($p_V = 0.29, 0.65$).  These exceptionally strong divergences satisfy the criteria required in high-order VHSs. The secondary high-order VHSs thus can be used as a signature of excitonic instabilities in materials. This is plausible since the optical spectra (i.e. the joint DOS) of many semiconductors and insulators are dominated by prominent VHSs and an analysis of their associated dispersion geometries would ease the identification of excitonic states \cite{RSMdd, Liang4,JCP}.  
 
{\it High-order VHSs in Bosonic systems.$-$} Since VHSs are crucial in the bosonic systems, one can ask the question if those VHSs can be of high-order. Here we demonstrate that the Bosonic VHSs carry features similar to the secondary VHSs discussed above. Reference~\onlinecite{MBMB} introduced the idea of a susceptibility density of states (SDOS) and showed its usefulness in mode-coupling theory and as a map of Fermi surface nesting. There has been recent interest in interpreting this susceptibility as a bosonic Green's function for electron-hole pairs\cite{Bosemet,spinon}.  We therefore consider $\chi_0(q,\omega)=\frac{1}{(\omega -\omega_q+i\gamma_q)}-\frac{1}{(\omega+\omega_q-i\gamma_q)}$, where $\omega_q$ is a bosonic (electron-hole) frequency and $\gamma_q$ a damping rate. For $\omega\rightarrow 0$ and $\gamma_q\rightarrow 0$, $\chi_0^{-1}=-\omega_q/2$ which gives the bosonic DOS up to a factor of 2, Fig.~\ref{fig:3c}. The dispersion in Fig.~\ref{fig:3c} looks unusual because the electronic susceptibility contains nonanalytic features (at $T=0$) due to Fermi surface nesting, which also show up in Kohn anomalies of phonons \cite{RSMch}.  Interestingly, we find that both the Mexican hat and drumhead (flat-band) dispersions exist in Bosonic systems and give rise to the high-order VHSs similar to shown in Fig.~\ref{fig:3b} of the electronic case (Fig.~2 of Ref.~\onlinecite{MBMB}). 

The bosonic high-order VHSs are particularly appealing since they make the transition from commensurate AFM to incommensurate spin-density wave (SDW) highly anomalous. At the crossover point, any signs of the electronic order are lost, leading to an emergent spin-glass phase. Since the AFM corresponds to what one expects of a Hubbard model (insensitive to shape of the Fermi surface) while the SDW is driven by Fermi surface nesting, it is appropriate to call this a Mott-Slater transition \cite{MBMB}, and the emergent spin-glass phase suggests why it is so hard to explain cuprate superconductivity starting from the undoped insulator. It should be noted that a very similar commensurate-incommensurate transition with hints of emergent spin-glass behavior at the crossover has been observed for the three-dimensional Hubbard model along with the finite-temperature phase transitions \cite{KohnAn}.

Insight into the bosonic ring and drumhead dispersions can be gained from the phononic dispersions. The electrons can be considered as moving in a quasi-static potential generated by the phonons, and a phonon soft-mode introduces a new component to the potential. The ring dispersion thus causes the electrons to move in a Mexican hat dispersion which is a signature of the Jahn-Teller effect. For phonons, the Mexican hat dispersion is typically connected to a point of conical intersection, where several phonon modes are degenerate. This point typically lies above the brim of the Mexican hat, signaling that the high-symmetry point is unstable.  Notably, the resulting strong electron-phonon coupling leads to highly exotic physics, including breakdown of the Born-Oppenheimer approximation and, possibly, time crystals \cite{Bob1993}.

{\it Discussion.$-$}
We have provided a comprehensive analysis of high-order VHSs in the cuprates and unveiled their relationship with the electronic dimensionality of the system. We now comment further on the varied roles these VHSs can play in cuprate physics. The inter-VHS nesting is responsible for the strong AFM effects relevant for Mott physics in the cuprates (Section III and SM II). A bosonic high-order VHS controls the transition from commensurate $(\pi,\pi)$ AFM to incommensurate SDW order, passing through an emergent spin-glass phase \cite{MBMB}. Since the SDW is sensitive to Fermi surface nesting, this may be considered as a Mott-Slater transition. The AFM quantum critical point as a function of doping seems to be connected with the crossover from $(\pi,\pi)$ VHSs to $\Gamma$-centered VHSs and this competition between the two VHS divergences may lead to the closing of the superconducting dome. For example, if we postulate that only the $(\pi,\pi)$ instability couples to superconductivity, then the following scenario emerges. First, in LSCO, $|t'|$ is small, and the $(\pi,\pi)$ instability dominates at all dopings, so $T_c$ is maximum near $x_{VHS}$. However, for most other cuprates $|t'|$ is larger, and there is a doping $x_{cross}<x_{VHS}$ where the dominant instability crosses over from $(\pi,\pi)$ to $\Gamma$. In this case, $T_c$ should maximize near $x_{cross}$ and decrease with larger doping. This scenario provides a good description of the cuprates. For example, in Bi2212 $T_c$ is maximal near $x_{cross}$, and $\rightarrow 0$ near $x_{VHS}$\cite{Zhou}, where the pairing strength also vanishes \cite{TallStorey}. 

Competition between inter-VHS scattering (which leads to near-$(\pi.\pi)$ AFM or SDW order)  and intra-VHS scattering (which favors longer-wavelength CDW order) can lead to a particularly intriguing form of intertwined order.  High -$T_c$ superconductivity could arise simply by tipping the balance between the competing orders, or it could actually benefit from the proximity of the strongly correlated and frustrated Mott phase with the bad metal Slater phase\cite{ABB}.




After years of debate on the significance of the VHSs in the cuprates, it is exciting to realize that they can be significantly more singular than previously imagined. The correlation between higher VHSs and higher superconducting $T_c$ we delineate here shows that high-order VHSs play a crucial role in high-$T_c$ cuprate superconductors. Understanding their extensive role will require answers to several issues, including understanding the dual role of a VHS, in both increasing correlations via the peak in the DOS/susceptibility, and decreasing correlations by enhancing dielectric screening. One also must understand the competition between intra- and inter-VHS coupling, and the significance of $x_{cross}$, as well as the role of secondary VHSs in driving/suppressing further instabilities. All of these studies will require accurate, material-specific calculations of electronic susceptibility, without introducing artificial broadening. The bosonic high-order VHSs may have important relevance to Bose metals \cite{Bosemet} and spinon bands \cite{spinon}. There is an ongoing search for exotic phase transitions that do not fit into the conventional Moriya-Hertz-Millis model of quantum criticality, particularly when nontrivial emergent excitations arise near the quantum critical point \cite{Sachdev}. Thus, our finding that such an emergent phase can be driven by bosonic high-order VHSs may constitute the most direct evidence of the importance of high-order VHSs in correlated materials. 

\bibliographystyle{myprsty2017}
\bibliography{HighOrder}

\begin{thebibliography}{33}%
\makeatletter
\providecommand \@ifxundefined [1]{%
 \@ifx{#1\undefined}
}%
\providecommand \@ifnum [1]{%
 \ifnum #1\expandafter \@firstoftwo
 \else \expandafter \@secondoftwo
 \fi
}%
\providecommand \@ifx [1]{%
 \ifx #1\expandafter \@firstoftwo
 \else \expandafter \@secondoftwo
 \fi
}%
\providecommand \natexlab [1]{#1}%
\providecommand \enquote  [1]{``#1''}%
\providecommand \bibnamefont  [1]{#1}%
\providecommand \bibfnamefont [1]{#1}%
\providecommand \citenamefont [1]{#1}%
\providecommand \href@noop [0]{\@secondoftwo}%
\providecommand \href [0]{\begingroup \@sanitize@url \@href}%
\providecommand \@href[1]{\@@startlink{#1}\@@href}%
\providecommand \@@href[1]{\endgroup#1\@@endlink}%
\providecommand \@sanitize@url [0]{\catcode `\\12\catcode `\$12\catcode
  `\&12\catcode `\#12\catcode `\^12\catcode `\_12\catcode `\%12\relax}%
\providecommand \@@startlink[1]{}%
\providecommand \@@endlink[0]{}%
\providecommand \url  [0]{\begingroup\@sanitize@url \@url }%
\providecommand \@url [1]{\endgroup\@href {#1}{\urlprefix }}%
\providecommand \urlprefix  [0]{URL }%
\providecommand \Eprint [0]{\href }%
\providecommand \doibase [0]{http://dx.doi.org/}%
\providecommand \selectlanguage [0]{\@gobble}%
\providecommand \bibinfo  [0]{\@secondoftwo}%
\providecommand \bibfield  [0]{\@secondoftwo}%
\providecommand \translation [1]{[#1]}%
\providecommand \BibitemOpen [0]{}%
\providecommand \bibitemStop [0]{}%
\providecommand \bibitemNoStop [0]{.\EOS\space}%
\providecommand \EOS [0]{\spacefactor3000\relax}%
\providecommand \BibitemShut  [1]{\csname bibitem#1\endcsname}%
\let\auto@bib@innerbib\@empty
\bibitem [{\citenamefont {Van~Hove}(1953)}]{VHS}%
  \BibitemOpen
  \bibfield  {author} {\bibinfo {author} {\bibfnamefont {L\'eon}\ \bibnamefont
  {Van~Hove}},\ }\bibfield  {title} {\enquote {\bibinfo {title} {The occurrence
  of singularities in the elastic frequency distribution of a crystal},}\
  }\href {\doibase 10.1103/PhysRev.89.1189} {\bibfield  {journal} {\bibinfo
  {journal} {Phys. Rev.}\ }\textbf {\bibinfo {volume} {89}},\ \bibinfo {pages}
  {1189--1193} (\bibinfo {year} {1953})}\BibitemShut {NoStop}%
\bibitem [{\citenamefont {Yuan}\ \emph {et~al.}(2019)\citenamefont {Yuan},
  \citenamefont {Isobe},\ and\ \citenamefont {Fu}}]{Liang}%
  \BibitemOpen
  \bibfield  {author} {\bibinfo {author} {\bibfnamefont {Noah F.~Q.}\
  \bibnamefont {Yuan}}, \bibinfo {author} {\bibfnamefont {Hiroki}\ \bibnamefont
  {Isobe}}, \ and\ \bibinfo {author} {\bibfnamefont {Liang}\ \bibnamefont
  {Fu}},\ }\bibfield  {title} {\enquote {\bibinfo {title} {Magic of high-order
  van hove singularity},}\ }\href {\doibase 10.1038/s41467-019-13670-9}
  {\bibfield  {journal} {\bibinfo  {journal} {Nature Communications}\ }\textbf
  {\bibinfo {volume} {10}},\ \bibinfo {pages} {5769} (\bibinfo {year}
  {2019})}\BibitemShut {NoStop}%
\bibitem [{\citenamefont {Yuan}\ and\ \citenamefont {Fu}(2020)}]{Liang3}%
  \BibitemOpen
  \bibfield  {author} {\bibinfo {author} {\bibfnamefont {Noah F.~Q.}\
  \bibnamefont {Yuan}}\ and\ \bibinfo {author} {\bibfnamefont {Liang}\
  \bibnamefont {Fu}},\ }\bibfield  {title} {\enquote {\bibinfo {title}
  {Classification of critical points in energy bands based on topology,
  scaling, and symmetry},}\ }\href {\doibase 10.1103/PhysRevB.101.125120}
  {\bibfield  {journal} {\bibinfo  {journal} {Phys. Rev. B}\ }\textbf {\bibinfo
  {volume} {101}},\ \bibinfo {pages} {125120} (\bibinfo {year}
  {2020})}\BibitemShut {NoStop}%
\bibitem [{\citenamefont {Markiewicz}\ \emph {et~al.}(2017)\citenamefont
  {Markiewicz}, \citenamefont {Buda}, \citenamefont {Mistark}, \citenamefont
  {Lane},\ and\ \citenamefont {Bansil}}]{MBMB}%
  \BibitemOpen
  \bibfield  {author} {\bibinfo {author} {\bibfnamefont {R.~S.}\ \bibnamefont
  {Markiewicz}}, \bibinfo {author} {\bibfnamefont {I.~G.}\ \bibnamefont
  {Buda}}, \bibinfo {author} {\bibfnamefont {P.}~\bibnamefont {Mistark}},
  \bibinfo {author} {\bibfnamefont {C.}~\bibnamefont {Lane}}, \ and\ \bibinfo
  {author} {\bibfnamefont {A.}~\bibnamefont {Bansil}},\ }\bibfield  {title}
  {\enquote {\bibinfo {title} {Entropic origin of pseudogap physics and a
  mott-slater transition in cuprates},}\ }\href {\doibase 10.1038/srep44008}
  {\bibfield  {journal} {\bibinfo  {journal} {Scientific Reports}\ }\textbf
  {\bibinfo {volume} {7}},\ \bibinfo {pages} {44008} (\bibinfo {year}
  {2017})}\BibitemShut {NoStop}%
\bibitem [{\citenamefont {Souza}\ and\ \citenamefont
  {Marsiglio}(2017)}]{Marsig}%
  \BibitemOpen
  \bibfield  {author} {\bibinfo {author} {\bibfnamefont {Thiago X.~R.}\
  \bibnamefont {Souza}}\ and\ \bibinfo {author} {\bibfnamefont
  {F.}~\bibnamefont {Marsiglio}},\ }\bibfield  {title} {\enquote {\bibinfo
  {title} {The possible role of van hove singularities in the high $t_c$ of
  superconducting h$_3$s},}\ }\href {\doibase 10.1142/S0217979217450035}
  {\bibfield  {journal} {\bibinfo  {journal} {International Journal of Modern
  Physics B}\ }\textbf {\bibinfo {volume} {31}},\ \bibinfo {pages} {1745003}
  (\bibinfo {year} {2017})}\BibitemShut {NoStop}%
\bibitem [{\citenamefont {Isobe}\ and\ \citenamefont {Fu}(2019)}]{Liang2}%
  \BibitemOpen
  \bibfield  {author} {\bibinfo {author} {\bibfnamefont {Hiroki}\ \bibnamefont
  {Isobe}}\ and\ \bibinfo {author} {\bibfnamefont {Liang}\ \bibnamefont {Fu}},\
  }\bibfield  {title} {\enquote {\bibinfo {title} {Supermetal},}\ }\href
  {\doibase 10.1103/PhysRevResearch.1.033206} {\bibfield  {journal} {\bibinfo
  {journal} {Phys. Rev. Research}\ }\textbf {\bibinfo {volume} {1}},\ \bibinfo
  {pages} {033206} (\bibinfo {year} {2019})}\BibitemShut {NoStop}%
\bibitem [{\citenamefont {Kerelsky}\ \emph {et~al.}(2019)\citenamefont
  {Kerelsky}, \citenamefont {McGilly}, \citenamefont {Kennes}, \citenamefont
  {Xian}, \citenamefont {Yankowitz}, \citenamefont {Chen}, \citenamefont
  {Watanabe}, \citenamefont {Taniguchi}, \citenamefont {Hone}, \citenamefont
  {Dean}, \citenamefont {Rubio},\ and\ \citenamefont {Pasupathy}}]{Tgraphene}%
  \BibitemOpen
  \bibfield  {author} {\bibinfo {author} {\bibfnamefont {Alexander}\
  \bibnamefont {Kerelsky}}, \bibinfo {author} {\bibfnamefont {Leo~J.}\
  \bibnamefont {McGilly}}, \bibinfo {author} {\bibfnamefont {Dante~M.}\
  \bibnamefont {Kennes}}, \bibinfo {author} {\bibfnamefont {Lede}\ \bibnamefont
  {Xian}}, \bibinfo {author} {\bibfnamefont {Matthew}\ \bibnamefont
  {Yankowitz}}, \bibinfo {author} {\bibfnamefont {Shaowen}\ \bibnamefont
  {Chen}}, \bibinfo {author} {\bibfnamefont {K.}~\bibnamefont {Watanabe}},
  \bibinfo {author} {\bibfnamefont {T.}~\bibnamefont {Taniguchi}}, \bibinfo
  {author} {\bibfnamefont {James}\ \bibnamefont {Hone}}, \bibinfo {author}
  {\bibfnamefont {Cory}\ \bibnamefont {Dean}}, \bibinfo {author} {\bibfnamefont
  {Angel}\ \bibnamefont {Rubio}}, \ and\ \bibinfo {author} {\bibfnamefont
  {Abhay~N.}\ \bibnamefont {Pasupathy}},\ }\bibfield  {title} {\enquote
  {\bibinfo {title} {Maximized electron interactions at the magic angle in
  twisted bilayer graphene},}\ }\href {\doibase 10.1038/s41586-019-1431-9}
  {\bibfield  {journal} {\bibinfo  {journal} {Nature}\ }\textbf {\bibinfo
  {volume} {572}},\ \bibinfo {pages} {95--100} (\bibinfo {year}
  {2019})}\BibitemShut {NoStop}%
\bibitem [{\citenamefont {Efremov}\ \emph {et~al.}(2019)\citenamefont
  {Efremov}, \citenamefont {Shtyk}, \citenamefont {Rost}, \citenamefont
  {Chamon}, \citenamefont {Mackenzie},\ and\ \citenamefont
  {Betouras}}]{SRO327}%
  \BibitemOpen
  \bibfield  {author} {\bibinfo {author} {\bibfnamefont {Dmitry~V.}\
  \bibnamefont {Efremov}}, \bibinfo {author} {\bibfnamefont {Alex}\
  \bibnamefont {Shtyk}}, \bibinfo {author} {\bibfnamefont {Andreas~W.}\
  \bibnamefont {Rost}}, \bibinfo {author} {\bibfnamefont {Claudio}\
  \bibnamefont {Chamon}}, \bibinfo {author} {\bibfnamefont {Andrew~P.}\
  \bibnamefont {Mackenzie}}, \ and\ \bibinfo {author} {\bibfnamefont
  {Joseph~J.}\ \bibnamefont {Betouras}},\ }\bibfield  {title} {\enquote
  {\bibinfo {title} {Multicritical fermi surface topological transitions},}\
  }\href {\doibase 10.1103/PhysRevLett.123.207202} {\bibfield  {journal}
  {\bibinfo  {journal} {Phys. Rev. Lett.}\ }\textbf {\bibinfo {volume} {123}},\
  \bibinfo {pages} {207202} (\bibinfo {year} {2019})}\BibitemShut {NoStop}%
\bibitem [{\citenamefont {Hirsch}\ and\ \citenamefont
  {Scalapino}(1986)}]{HirSc}%
  \BibitemOpen
  \bibfield  {author} {\bibinfo {author} {\bibfnamefont {J.~E.}\ \bibnamefont
  {Hirsch}}\ and\ \bibinfo {author} {\bibfnamefont {D.~J.}\ \bibnamefont
  {Scalapino}},\ }\bibfield  {title} {\enquote {\bibinfo {title} {Enhanced
  superconductivity in quasi two-dimensional systems},}\ }\href {\doibase
  10.1103/PhysRevLett.56.2732} {\bibfield  {journal} {\bibinfo  {journal}
  {Phys. Rev. Lett.}\ }\textbf {\bibinfo {volume} {56}},\ \bibinfo {pages}
  {2732--2735} (\bibinfo {year} {1986})}\BibitemShut {NoStop}%
\bibitem [{\citenamefont {Michon}\ \emph {et~al.}(2019)\citenamefont {Michon},
  \citenamefont {Girod}, \citenamefont {Badoux}, \citenamefont {Ka{\v c}mar{\v
  c}{\'\i}k}, \citenamefont {Ma}, \citenamefont {Dragomir}, \citenamefont
  {Dabkowska}, \citenamefont {Gaulin}, \citenamefont {Zhou}, \citenamefont
  {Pyon}, \citenamefont {Takayama}, \citenamefont {Takagi}, \citenamefont
  {Verret}, \citenamefont {Doiron-Leyraud}, \citenamefont {Marcenat},
  \citenamefont {Taillefer},\ and\ \citenamefont {Klein}}]{Michon}%
  \BibitemOpen
  \bibfield  {author} {\bibinfo {author} {\bibfnamefont {B.}~\bibnamefont
  {Michon}}, \bibinfo {author} {\bibfnamefont {C.}~\bibnamefont {Girod}},
  \bibinfo {author} {\bibfnamefont {S.}~\bibnamefont {Badoux}}, \bibinfo
  {author} {\bibfnamefont {J.}~\bibnamefont {Ka{\v c}mar{\v c}{\'\i}k}},
  \bibinfo {author} {\bibfnamefont {Q.}~\bibnamefont {Ma}}, \bibinfo {author}
  {\bibfnamefont {M.}~\bibnamefont {Dragomir}}, \bibinfo {author}
  {\bibfnamefont {H.~A.}\ \bibnamefont {Dabkowska}}, \bibinfo {author}
  {\bibfnamefont {B.~D.}\ \bibnamefont {Gaulin}}, \bibinfo {author}
  {\bibfnamefont {J.~S.}\ \bibnamefont {Zhou}}, \bibinfo {author}
  {\bibfnamefont {S.}~\bibnamefont {Pyon}}, \bibinfo {author} {\bibfnamefont
  {T.}~\bibnamefont {Takayama}}, \bibinfo {author} {\bibfnamefont
  {H.}~\bibnamefont {Takagi}}, \bibinfo {author} {\bibfnamefont
  {S.}~\bibnamefont {Verret}}, \bibinfo {author} {\bibfnamefont
  {N.}~\bibnamefont {Doiron-Leyraud}}, \bibinfo {author} {\bibfnamefont
  {C.}~\bibnamefont {Marcenat}}, \bibinfo {author} {\bibfnamefont
  {L.}~\bibnamefont {Taillefer}}, \ and\ \bibinfo {author} {\bibfnamefont
  {T.}~\bibnamefont {Klein}},\ }\bibfield  {title} {\enquote {\bibinfo {title}
  {Thermodynamic signatures of quantum criticality in cuprate
  superconductors},}\ }\href {\doibase 10.1038/s41586-019-0932-x} {\bibfield
  {journal} {\bibinfo  {journal} {Nature}\ }\textbf {\bibinfo {volume} {567}},\
  \bibinfo {pages} {218--222} (\bibinfo {year} {2019})}\BibitemShut {NoStop}%
\bibitem [{\citenamefont {Markiewicz}(2004)}]{RM70}%
  \BibitemOpen
  \bibfield  {author} {\bibinfo {author} {\bibfnamefont {R.~S.}\ \bibnamefont
  {Markiewicz}},\ }\bibfield  {title} {\enquote {\bibinfo {title}
  {Mode-coupling model of mott gap collapse in the cuprates: Natural phase
  boundary for quantum critical points},}\ }\href {\doibase
  10.1103/PhysRevB.70.174518} {\bibfield  {journal} {\bibinfo  {journal} {Phys.
  Rev. B}\ }\textbf {\bibinfo {volume} {70}},\ \bibinfo {pages} {174518}
  (\bibinfo {year} {2004})}\BibitemShut {NoStop}%
\bibitem [{\citenamefont {Lizaire}\ \emph {et~al.}(2020)\citenamefont
  {Lizaire}, \citenamefont {Legros}, \citenamefont {Gourgout}, \citenamefont
  {Benhabib}, \citenamefont {Badoux}, \citenamefont {Laliberte}, \citenamefont
  {Boulanger}, \citenamefont {Ataei}, \citenamefont {Grissonnanche},
  \citenamefont {LeBoeuf}, \citenamefont {Licciardello}, \citenamefont
  {Wiedmann}, \citenamefont {S.~Ono}, \citenamefont {Zheng}, \citenamefont
  {Doiron-Leyraud}, \citenamefont {Proust},\ and\ \citenamefont
  {Taillefer}}]{Lizaire2020}%
  \BibitemOpen
  \bibfield  {author} {\bibinfo {author} {\bibfnamefont {M.}~\bibnamefont
  {Lizaire}}, \bibinfo {author} {\bibfnamefont {A.}~\bibnamefont {Legros}},
  \bibinfo {author} {\bibfnamefont {A.}~\bibnamefont {Gourgout}}, \bibinfo
  {author} {\bibfnamefont {S.}~\bibnamefont {Benhabib}}, \bibinfo {author}
  {\bibfnamefont {S.}~\bibnamefont {Badoux}}, \bibinfo {author} {\bibfnamefont
  {F.}~\bibnamefont {Laliberte}}, \bibinfo {author} {\bibfnamefont {M.~E.}\
  \bibnamefont {Boulanger}}, \bibinfo {author} {\bibfnamefont {A.}~\bibnamefont
  {Ataei}}, \bibinfo {author} {\bibfnamefont {G.}~\bibnamefont
  {Grissonnanche}}, \bibinfo {author} {\bibfnamefont {D.}~\bibnamefont
  {LeBoeuf}}, \bibinfo {author} {\bibfnamefont {S.}~\bibnamefont
  {Licciardello}}, \bibinfo {author} {\bibfnamefont {S.}~\bibnamefont
  {Wiedmann}}, \bibinfo {author} {\bibfnamefont {S.~Kawasaki}\ \bibnamefont
  {S.~Ono}}, \bibinfo {author} {\bibfnamefont {G.~Q.}\ \bibnamefont {Zheng}},
  \bibinfo {author} {\bibfnamefont {N.}~\bibnamefont {Doiron-Leyraud}},
  \bibinfo {author} {\bibfnamefont {C.}~\bibnamefont {Proust}}, \ and\ \bibinfo
  {author} {\bibfnamefont {L.}~\bibnamefont {Taillefer}},\ }\bibfield  {title}
  {\enquote {\bibinfo {title} {Transport signatures of the pseudogap critical
  point in the cuprate superconductor
  bi$_2$sr$_{2-x}$la$_x$cuo$_{6+\delta}$},}\ }\href
  {https://arxiv.org/abs/2008.13692} {\bibfield  {journal} {\bibinfo  {journal}
  {ArXiv:2008.13692}\ } (\bibinfo {year} {2020})}\BibitemShut {NoStop}%
\bibitem [{\citenamefont {Pavarini}\ \emph {et~al.}(2001)\citenamefont
  {Pavarini}, \citenamefont {Dasgupta}, \citenamefont {Saha-Dasgupta},
  \citenamefont {Jepsen},\ and\ \citenamefont {Andersen}}]{PavOK}%
  \BibitemOpen
  \bibfield  {author} {\bibinfo {author} {\bibfnamefont {E.}~\bibnamefont
  {Pavarini}}, \bibinfo {author} {\bibfnamefont {I.}~\bibnamefont {Dasgupta}},
  \bibinfo {author} {\bibfnamefont {T.}~\bibnamefont {Saha-Dasgupta}}, \bibinfo
  {author} {\bibfnamefont {O.}~\bibnamefont {Jepsen}}, \ and\ \bibinfo {author}
  {\bibfnamefont {O.~K.}\ \bibnamefont {Andersen}},\ }\bibfield  {title}
  {\enquote {\bibinfo {title} {Band-structure trend in hole-doped cuprates and
  correlation with ${\mathit{t}}_{\mathit{c}\mathrm{max}}$},}\ }\href {\doibase
  10.1103/PhysRevLett.87.047003} {\bibfield  {journal} {\bibinfo  {journal}
  {Phys. Rev. Lett.}\ }\textbf {\bibinfo {volume} {87}},\ \bibinfo {pages}
  {047003} (\bibinfo {year} {2001})}\BibitemShut {NoStop}%
\bibitem [{\citenamefont {Andersen}\ \emph {et~al.}(1994)\citenamefont
  {Andersen}, \citenamefont {Jepsen}, \citenamefont {Liechtenstein},\ and\
  \citenamefont {Mazin}}]{Andex}%
  \BibitemOpen
  \bibfield  {author} {\bibinfo {author} {\bibfnamefont {O.~K.}\ \bibnamefont
  {Andersen}}, \bibinfo {author} {\bibfnamefont {O.}~\bibnamefont {Jepsen}},
  \bibinfo {author} {\bibfnamefont {A.~I.}\ \bibnamefont {Liechtenstein}}, \
  and\ \bibinfo {author} {\bibfnamefont {I.~I.}\ \bibnamefont {Mazin}},\
  }\bibfield  {title} {\enquote {\bibinfo {title} {Plane dimpling and
  saddle-point bifurcation in the band structures of optimally doped
  high-temperature superconductors: A tight-binding model},}\ }\href {\doibase
  10.1103/PhysRevB.49.4145} {\bibfield  {journal} {\bibinfo  {journal} {Phys.
  Rev. B}\ }\textbf {\bibinfo {volume} {49}},\ \bibinfo {pages} {4145--4157}
  (\bibinfo {year} {1994})}\BibitemShut {NoStop}%
\bibitem [{\citenamefont {Labbe}\ and\ \citenamefont {Friedel}(1966)}]{Labbe}%
  \BibitemOpen
  \bibfield  {author} {\bibinfo {author} {\bibfnamefont {J.}~\bibnamefont
  {Labbe}}\ and\ \bibinfo {author} {\bibfnamefont {J.}~\bibnamefont
  {Friedel}},\ }\bibfield  {title} {\enquote {\bibinfo {title} {Classification
  of critical points in energy bands based on topology, scaling, and
  symmetry},}\ }\href {\doibase
  https://doi.org/10.1051/jphys:01966002703-4015300} {\bibfield  {journal}
  {\bibinfo  {journal} {J. Phys. France}\ }\textbf {\bibinfo {volume} {27}},\
  \bibinfo {pages} {153 -- 165} (\bibinfo {year} {1966})}\BibitemShut {NoStop}%
\bibitem [{\citenamefont {Markiewicz}\ and\ \citenamefont
  {Vaughn}(1998)}]{Mikefest}%
  \BibitemOpen
  \bibfield  {author} {\bibinfo {author} {\bibfnamefont {R.~S.}\ \bibnamefont
  {Markiewicz}}\ and\ \bibinfo {author} {\bibfnamefont {M.~T.}\ \bibnamefont
  {Vaughn}},\ }\bibfield  {title} {\enquote {\bibinfo {title} {Classification
  of the van hove scenario as an so(8) spectrum-generating algebra},}\ }\href
  {\doibase 10.1103/PhysRevB.57.R14052} {\bibfield  {journal} {\bibinfo
  {journal} {Phys. Rev. B}\ }\textbf {\bibinfo {volume} {57}},\ \bibinfo
  {pages} {R14052} (\bibinfo {year} {1998})}\BibitemShut {NoStop}%
\bibitem [{\citenamefont {Markiewicz}(2000)}]{RSMstripe}%
  \BibitemOpen
  \bibfield  {author} {\bibinfo {author} {\bibfnamefont {R.~S.}\ \bibnamefont
  {Markiewicz}},\ }\bibfield  {title} {\enquote {\bibinfo {title} {Dispersion
  of ordered stripe phases in the cuprates},}\ }\href {\doibase
  10.1103/PhysRevB.62.1252} {\bibfield  {journal} {\bibinfo  {journal} {Phys.
  Rev. B}\ }\textbf {\bibinfo {volume} {62}},\ \bibinfo {pages} {1252--1269}
  (\bibinfo {year} {2000})}\BibitemShut {NoStop}%
\bibitem [{\citenamefont {Horio}\ \emph {et~al.}(2018)\citenamefont {Horio},
  \citenamefont {Hauser}, \citenamefont {Sassa}, \citenamefont {Mingazheva},
  \citenamefont {Sutter}, \citenamefont {Kramer}, \citenamefont {Cook},
  \citenamefont {Nocerino}, \citenamefont {Forslund}, \citenamefont
  {Tjernberg}, \citenamefont {Kobayashi}, \citenamefont {Chikina},
  \citenamefont {Schr\"oter}, \citenamefont {Krieger}, \citenamefont {Schmitt},
  \citenamefont {Strocov}, \citenamefont {Pyon}, \citenamefont {Takayama},
  \citenamefont {Takagi}, \citenamefont {Lipscombe}, \citenamefont {Hayden},
  \citenamefont {Ishikado}, \citenamefont {Eisaki}, \citenamefont {Neupert},
  \citenamefont {M\aa{}nsson}, \citenamefont {Matt},\ and\ \citenamefont
  {Chang}}]{Horio}%
  \BibitemOpen
  \bibfield  {author} {\bibinfo {author} {\bibfnamefont {M.}~\bibnamefont
  {Horio}}, \bibinfo {author} {\bibfnamefont {K.}~\bibnamefont {Hauser}},
  \bibinfo {author} {\bibfnamefont {Y.}~\bibnamefont {Sassa}}, \bibinfo
  {author} {\bibfnamefont {Z.}~\bibnamefont {Mingazheva}}, \bibinfo {author}
  {\bibfnamefont {D.}~\bibnamefont {Sutter}}, \bibinfo {author} {\bibfnamefont
  {K.}~\bibnamefont {Kramer}}, \bibinfo {author} {\bibfnamefont
  {A.}~\bibnamefont {Cook}}, \bibinfo {author} {\bibfnamefont {E.}~\bibnamefont
  {Nocerino}}, \bibinfo {author} {\bibfnamefont {O.~K.}\ \bibnamefont
  {Forslund}}, \bibinfo {author} {\bibfnamefont {O.}~\bibnamefont {Tjernberg}},
  \bibinfo {author} {\bibfnamefont {M.}~\bibnamefont {Kobayashi}}, \bibinfo
  {author} {\bibfnamefont {A.}~\bibnamefont {Chikina}}, \bibinfo {author}
  {\bibfnamefont {N.~B.~M.}\ \bibnamefont {Schr\"oter}}, \bibinfo {author}
  {\bibfnamefont {J.~A.}\ \bibnamefont {Krieger}}, \bibinfo {author}
  {\bibfnamefont {T.}~\bibnamefont {Schmitt}}, \bibinfo {author} {\bibfnamefont
  {V.~N.}\ \bibnamefont {Strocov}}, \bibinfo {author} {\bibfnamefont
  {S.}~\bibnamefont {Pyon}}, \bibinfo {author} {\bibfnamefont {T.}~\bibnamefont
  {Takayama}}, \bibinfo {author} {\bibfnamefont {H.}~\bibnamefont {Takagi}},
  \bibinfo {author} {\bibfnamefont {O.~J.}\ \bibnamefont {Lipscombe}}, \bibinfo
  {author} {\bibfnamefont {S.~M.}\ \bibnamefont {Hayden}}, \bibinfo {author}
  {\bibfnamefont {M.}~\bibnamefont {Ishikado}}, \bibinfo {author}
  {\bibfnamefont {H.}~\bibnamefont {Eisaki}}, \bibinfo {author} {\bibfnamefont
  {T.}~\bibnamefont {Neupert}}, \bibinfo {author} {\bibfnamefont
  {M.}~\bibnamefont {M\aa{}nsson}}, \bibinfo {author} {\bibfnamefont {C.~E.}\
  \bibnamefont {Matt}}, \ and\ \bibinfo {author} {\bibfnamefont
  {J.}~\bibnamefont {Chang}},\ }\bibfield  {title} {\enquote {\bibinfo {title}
  {Three-dimensional fermi surface of overdoped la-based cuprates},}\ }\href
  {\doibase 10.1103/PhysRevLett.121.077004} {\bibfield  {journal} {\bibinfo
  {journal} {Phys. Rev. Lett.}\ }\textbf {\bibinfo {volume} {121}},\ \bibinfo
  {pages} {077004} (\bibinfo {year} {2018})}\BibitemShut {NoStop}%
\bibitem [{\citenamefont {Singh}\ \emph {et~al.}(2017)\citenamefont {Singh},
  \citenamefont {Hsu}, \citenamefont {Tsai}, \citenamefont {Pereira},\ and\
  \citenamefont {Lin}}]{singh2017}%
  \BibitemOpen
  \bibfield  {author} {\bibinfo {author} {\bibfnamefont {Bahadur}\ \bibnamefont
  {Singh}}, \bibinfo {author} {\bibfnamefont {Chuang-Han}\ \bibnamefont {Hsu}},
  \bibinfo {author} {\bibfnamefont {Wei-Feng}\ \bibnamefont {Tsai}}, \bibinfo
  {author} {\bibfnamefont {Vitor~M.}\ \bibnamefont {Pereira}}, \ and\ \bibinfo
  {author} {\bibfnamefont {Hsin}\ \bibnamefont {Lin}},\ }\bibfield  {title}
  {\enquote {\bibinfo {title} {Stable charge density wave phase in a
  $1t--{\mathrm{tise}}_{2}$ monolayer},}\ }\href {\doibase
  10.1103/PhysRevB.95.245136} {\bibfield  {journal} {\bibinfo  {journal} {Phys.
  Rev. B}\ }\textbf {\bibinfo {volume} {95}},\ \bibinfo {pages} {245136}
  (\bibinfo {year} {2017})}\BibitemShut {NoStop}%
\bibitem [{\citenamefont {Bansil}\ \emph {et~al.}(2016)\citenamefont {Bansil},
  \citenamefont {Lin},\ and\ \citenamefont {Das}}]{Bansil2016}%
  \BibitemOpen
  \bibfield  {author} {\bibinfo {author} {\bibfnamefont {A.}~\bibnamefont
  {Bansil}}, \bibinfo {author} {\bibfnamefont {Hsin}\ \bibnamefont {Lin}}, \
  and\ \bibinfo {author} {\bibfnamefont {Tanmoy}\ \bibnamefont {Das}},\
  }\bibfield  {title} {\enquote {\bibinfo {title} {Colloquium: Topological band
  theory},}\ }\href {\doibase 10.1103/RevModPhys.88.021004} {\bibfield
  {journal} {\bibinfo  {journal} {Rev. Mod. Phys.}\ }\textbf {\bibinfo {volume}
  {88}},\ \bibinfo {pages} {021004} (\bibinfo {year} {2016})}\BibitemShut
  {NoStop}%
\bibitem [{\citenamefont {Overhauser}(1968)}]{Over}%
  \BibitemOpen
  \bibfield  {author} {\bibinfo {author} {\bibfnamefont {A.~W.}\ \bibnamefont
  {Overhauser}},\ }\bibfield  {title} {\enquote {\bibinfo {title} {Exchange and
  correlation instabilities of simple metals},}\ }\href {\doibase
  10.1103/PhysRev.167.691} {\bibfield  {journal} {\bibinfo  {journal} {Phys.
  Rev.}\ }\textbf {\bibinfo {volume} {167}},\ \bibinfo {pages} {691--698}
  (\bibinfo {year} {1968})}\BibitemShut {NoStop}%
\bibitem [{\citenamefont {Markiewicz}\ and\ \citenamefont
  {Bansil}(2018)}]{RSMdd}%
  \BibitemOpen
  \bibfield  {author} {\bibinfo {author} {\bibfnamefont {R.~S.}\ \bibnamefont
  {Markiewicz}}\ and\ \bibinfo {author} {\bibfnamefont {A.}~\bibnamefont
  {Bansil}},\ }\bibfield  {title} {\enquote {\bibinfo {title} {Excitonic
  insulators as a model of $d-d$ and mott transitions in strongly correlated
  materials},}\ }\href {https://arxiv.org/abs/1708.02270} {\bibfield  {journal}
  {\bibinfo  {journal} {ArXiv:1708.02270}\ } (\bibinfo {year}
  {2018})}\BibitemShut {NoStop}%
\bibitem [{\citenamefont {Fu}\ and\ \citenamefont {Bi}(2019)}]{Liang4}%
  \BibitemOpen
  \bibfield  {author} {\bibinfo {author} {\bibfnamefont {Liang}\ \bibnamefont
  {Fu}}\ and\ \bibinfo {author} {\bibfnamefont {Zhen}\ \bibnamefont {Bi}},\
  }\bibfield  {title} {\enquote {\bibinfo {title} {Excitonic density wave and
  spin-valley superfluid in bilayer transition metal dichalcogenide},}\ }\href
  {https://arxiv.org/abs/1911.04493} {\bibfield  {journal} {\bibinfo  {journal}
  {ArXiv:1911.04493}\ } (\bibinfo {year} {2019})}\BibitemShut {NoStop}%
\bibitem [{\citenamefont {Phillips}(1964)}]{JCP}%
  \BibitemOpen
  \bibfield  {author} {\bibinfo {author} {\bibfnamefont {J.~C.}\ \bibnamefont
  {Phillips}},\ }\bibfield  {title} {\enquote {\bibinfo {title} {Ultraviolet
  absorption of insulators. iii. fcc alkali halides},}\ }\href {\doibase
  10.1103/PhysRev.136.A1705} {\bibfield  {journal} {\bibinfo  {journal} {Phys.
  Rev.}\ }\textbf {\bibinfo {volume} {136}},\ \bibinfo {pages} {A1705}
  (\bibinfo {year} {1964})}\BibitemShut {NoStop}%
\bibitem [{\citenamefont {Sheng}\ \emph {et~al.}(2009)\citenamefont {Sheng},
  \citenamefont {Motrunich},\ and\ \citenamefont {Fisher}}]{Bosemet}%
  \BibitemOpen
  \bibfield  {author} {\bibinfo {author} {\bibfnamefont {D.~N.}\ \bibnamefont
  {Sheng}}, \bibinfo {author} {\bibfnamefont {Olexei~I.}\ \bibnamefont
  {Motrunich}}, \ and\ \bibinfo {author} {\bibfnamefont {Matthew P.~A.}\
  \bibnamefont {Fisher}},\ }\bibfield  {title} {\enquote {\bibinfo {title}
  {Spin bose-metal phase in a spin-$\frac{1}{2}$ model with ring exchange on a
  two-leg triangular strip},}\ }\href {\doibase 10.1103/PhysRevB.79.205112}
  {\bibfield  {journal} {\bibinfo  {journal} {Phys. Rev. B}\ }\textbf {\bibinfo
  {volume} {79}},\ \bibinfo {pages} {205112} (\bibinfo {year}
  {2009})}\BibitemShut {NoStop}%
\bibitem [{\citenamefont {Hu}\ \emph {et~al.}(2020)\citenamefont {Hu},
  \citenamefont {Zhang}, \citenamefont {Nevidomskyy}, \citenamefont {Dagotto},
  \citenamefont {Si},\ and\ \citenamefont {Lai}}]{spinon}%
  \BibitemOpen
  \bibfield  {author} {\bibinfo {author} {\bibfnamefont {Wen-Jun}\ \bibnamefont
  {Hu}}, \bibinfo {author} {\bibfnamefont {Yi}~\bibnamefont {Zhang}}, \bibinfo
  {author} {\bibfnamefont {Andriy~H.}\ \bibnamefont {Nevidomskyy}}, \bibinfo
  {author} {\bibfnamefont {Elbio}\ \bibnamefont {Dagotto}}, \bibinfo {author}
  {\bibfnamefont {Qimiao}\ \bibnamefont {Si}}, \ and\ \bibinfo {author}
  {\bibfnamefont {Hsin-Hua}\ \bibnamefont {Lai}},\ }\bibfield  {title}
  {\enquote {\bibinfo {title} {Fractionalized excitations revealed by
  entanglement entropy},}\ }\href {\doibase 10.1103/PhysRevLett.124.237201}
  {\bibfield  {journal} {\bibinfo  {journal} {Phys. Rev. Lett.}\ }\textbf
  {\bibinfo {volume} {124}},\ \bibinfo {pages} {237201} (\bibinfo {year}
  {2020})}\BibitemShut {NoStop}%
\bibitem [{\citenamefont {Markiewicz}\ \emph {et~al.}(2015)\citenamefont
  {Markiewicz}, \citenamefont {Seibold}, \citenamefont {Lorenzana},\ and\
  \citenamefont {Bansil}}]{RSMch}%
  \BibitemOpen
  \bibfield  {author} {\bibinfo {author} {\bibfnamefont {R.~S.}\ \bibnamefont
  {Markiewicz}}, \bibinfo {author} {\bibfnamefont {G.}~\bibnamefont {Seibold}},
  \bibinfo {author} {\bibfnamefont {J.}~\bibnamefont {Lorenzana}}, \ and\
  \bibinfo {author} {\bibfnamefont {A.}~\bibnamefont {Bansil}},\ }\bibfield
  {title} {\enquote {\bibinfo {title} {Gutzwiller charge phase diagram of
  cuprates, including electron{\textendash}phonon coupling effects},}\ }\href
  {\doibase 10.1088/1367-2630/17/2/023074} {\bibfield  {journal} {\bibinfo
  {journal} {New Journal of Physics}\ }\textbf {\bibinfo {volume} {17}},\
  \bibinfo {pages} {023074} (\bibinfo {year} {2015})}\BibitemShut {NoStop}%
\bibitem [{\citenamefont {Sch\"afer}\ \emph {et~al.}(2017)\citenamefont
  {Sch\"afer}, \citenamefont {Katanin}, \citenamefont {Held},\ and\
  \citenamefont {Toschi}}]{KohnAn}%
  \BibitemOpen
  \bibfield  {author} {\bibinfo {author} {\bibfnamefont {T.}~\bibnamefont
  {Sch\"afer}}, \bibinfo {author} {\bibfnamefont {A.~A.}\ \bibnamefont
  {Katanin}}, \bibinfo {author} {\bibfnamefont {K.}~\bibnamefont {Held}}, \
  and\ \bibinfo {author} {\bibfnamefont {A.}~\bibnamefont {Toschi}},\
  }\bibfield  {title} {\enquote {\bibinfo {title} {Interplay of correlations
  and kohn anomalies in three dimensions: Quantum criticality with a twist},}\
  }\href {\doibase 10.1103/PhysRevLett.119.046402} {\bibfield  {journal}
  {\bibinfo  {journal} {Phys. Rev. Lett.}\ }\textbf {\bibinfo {volume} {119}},\
  \bibinfo {pages} {046402} (\bibinfo {year} {2017})}\BibitemShut {NoStop}%
\bibitem [{\citenamefont {Markiewicz}(1993)}]{Bob1993}%
  \BibitemOpen
  \bibfield  {author} {\bibinfo {author} {\bibfnamefont {R.~S.}\ \bibnamefont
  {Markiewicz}},\ }\bibfield  {title} {\enquote {\bibinfo {title} {Van hove
  exciton-cageons and high-tc superconductivity: Viiid. solitons and nonlinear
  dynamics},}\ }\href {\doibase https://doi.org/10.1016/0921-4534(93)90030-T}
  {\bibfield  {journal} {\bibinfo  {journal} {Physica C: Superconductivity}\
  }\textbf {\bibinfo {volume} {210}},\ \bibinfo {pages} {264} (\bibinfo {year}
  {1993})}\BibitemShut {NoStop}%
\bibitem [{\citenamefont {Ding}\ \emph {et~al.}(2019)\citenamefont {Ding},
  \citenamefont {Zhao}, \citenamefont {Yan}, \citenamefont {Gao}, \citenamefont
  {Liu}, \citenamefont {Hu}, \citenamefont {Huang}, \citenamefont {Li},
  \citenamefont {Xu}, \citenamefont {Cai}, \citenamefont {Rong}, \citenamefont
  {Wu}, \citenamefont {Song}, \citenamefont {Zhou}, \citenamefont {Dong},
  \citenamefont {Liu}, \citenamefont {Wang}, \citenamefont {Zhang},
  \citenamefont {Wang}, \citenamefont {Zhang}, \citenamefont {Yang},
  \citenamefont {Peng}, \citenamefont {Xu}, \citenamefont {Chen},\ and\
  \citenamefont {Zhou}}]{Zhou}%
  \BibitemOpen
  \bibfield  {author} {\bibinfo {author} {\bibfnamefont {Ying}\ \bibnamefont
  {Ding}}, \bibinfo {author} {\bibfnamefont {Lin}\ \bibnamefont {Zhao}},
  \bibinfo {author} {\bibfnamefont {Hongtao}\ \bibnamefont {Yan}}, \bibinfo
  {author} {\bibfnamefont {Qiang}\ \bibnamefont {Gao}}, \bibinfo {author}
  {\bibfnamefont {Jing}\ \bibnamefont {Liu}}, \bibinfo {author} {\bibfnamefont
  {Cheng}\ \bibnamefont {Hu}}, \bibinfo {author} {\bibfnamefont {Jianwei}\
  \bibnamefont {Huang}}, \bibinfo {author} {\bibfnamefont {Cong}\ \bibnamefont
  {Li}}, \bibinfo {author} {\bibfnamefont {Yu}~\bibnamefont {Xu}}, \bibinfo
  {author} {\bibfnamefont {Yongqing}\ \bibnamefont {Cai}}, \bibinfo {author}
  {\bibfnamefont {Hongtao}\ \bibnamefont {Rong}}, \bibinfo {author}
  {\bibfnamefont {Dingsong}\ \bibnamefont {Wu}}, \bibinfo {author}
  {\bibfnamefont {Chunyao}\ \bibnamefont {Song}}, \bibinfo {author}
  {\bibfnamefont {Huaxue}\ \bibnamefont {Zhou}}, \bibinfo {author}
  {\bibfnamefont {Xiaoli}\ \bibnamefont {Dong}}, \bibinfo {author}
  {\bibfnamefont {Guodong}\ \bibnamefont {Liu}}, \bibinfo {author}
  {\bibfnamefont {Qingyan}\ \bibnamefont {Wang}}, \bibinfo {author}
  {\bibfnamefont {Shenjin}\ \bibnamefont {Zhang}}, \bibinfo {author}
  {\bibfnamefont {Zhimin}\ \bibnamefont {Wang}}, \bibinfo {author}
  {\bibfnamefont {Fengfeng}\ \bibnamefont {Zhang}}, \bibinfo {author}
  {\bibfnamefont {Feng}\ \bibnamefont {Yang}}, \bibinfo {author} {\bibfnamefont
  {Qinjun}\ \bibnamefont {Peng}}, \bibinfo {author} {\bibfnamefont {Zuyan}\
  \bibnamefont {Xu}}, \bibinfo {author} {\bibfnamefont {Chuangtian}\
  \bibnamefont {Chen}}, \ and\ \bibinfo {author} {\bibfnamefont {X.~J.}\
  \bibnamefont {Zhou}},\ }\bibfield  {title} {\enquote {\bibinfo {title}
  {Disappearance of superconductivity and a concomitant lifshitz transition in
  heavily overdoped bi$_2$sr$_2$cuo$_6$ superconductor revealed by
  angle-resolved photoemission spectroscopy},}\ }\href {\doibase
  10.1088/0256-307X/36/1/017402} {\bibfield  {journal} {\bibinfo  {journal}
  {Chin. Phys. Lett.}\ }\textbf {\bibinfo {volume} {36}},\ \bibinfo {pages}
  {017402} (\bibinfo {year} {2019})}\BibitemShut {NoStop}%
\bibitem [{\citenamefont {Storey}\ \emph {et~al.}(2007)\citenamefont {Storey},
  \citenamefont {Tallon},\ and\ \citenamefont {Williams}}]{TallStorey}%
  \BibitemOpen
  \bibfield  {author} {\bibinfo {author} {\bibfnamefont {J.~G.}\ \bibnamefont
  {Storey}}, \bibinfo {author} {\bibfnamefont {J.~L.}\ \bibnamefont {Tallon}},
  \ and\ \bibinfo {author} {\bibfnamefont {G.~V.~M.}\ \bibnamefont
  {Williams}},\ }\bibfield  {title} {\enquote {\bibinfo {title} {Saddle-point
  van hove singularity and the phase diagram of high-${T}_{c}$ cuprates},}\
  }\href {\doibase 10.1103/PhysRevB.76.174522} {\bibfield  {journal} {\bibinfo
  {journal} {Phys. Rev. B}\ }\textbf {\bibinfo {volume} {76}},\ \bibinfo
  {pages} {174522} (\bibinfo {year} {2007})}\BibitemShut {NoStop}%
\bibitem [{\citenamefont {Allender}\ \emph {et~al.}(1974)\citenamefont
  {Allender}, \citenamefont {Bray},\ and\ \citenamefont {Bardeen}}]{ABB}%
  \BibitemOpen
  \bibfield  {author} {\bibinfo {author} {\bibfnamefont {David}\ \bibnamefont
  {Allender}}, \bibinfo {author} {\bibfnamefont {J.~W.}\ \bibnamefont {Bray}},
  \ and\ \bibinfo {author} {\bibfnamefont {John}\ \bibnamefont {Bardeen}},\
  }\bibfield  {title} {\enquote {\bibinfo {title} {Theory of fluctuation
  superconductivity from electron-phonon interactions in pseudo-one-dimensional
  systems},}\ }\href {\doibase 10.1103/PhysRevB.9.119} {\bibfield  {journal}
  {\bibinfo  {journal} {Phys. Rev. B}\ }\textbf {\bibinfo {volume} {9}},\
  \bibinfo {pages} {119--129} (\bibinfo {year} {1974})}\BibitemShut {NoStop}%
\bibitem [{\citenamefont {Sachdev}(2000)}]{Sachdev}%
  \BibitemOpen
  \bibfield  {author} {\bibinfo {author} {\bibfnamefont {Subir}\ \bibnamefont
  {Sachdev}},\ }\bibfield  {title} {\enquote {\bibinfo {title} {Quantum
  criticality: Competing ground states in low dimensions},}\ }\href {\doibase
  10.1126/science.288.5465.475} {\bibfield  {journal} {\bibinfo  {journal}
  {Science}\ }\textbf {\bibinfo {volume} {288}},\ \bibinfo {pages} {475}
  (\bibinfo {year} {2000})}\BibitemShut {NoStop}%
\end{thebibliography}%

\section*{Acknowledgements}     
This work is supported by the US Department of Energy, Office of Science, Basic Energy Sciences grant number DE-FG02-07ER46352, and benefited from Northeastern University's Advanced Scientific Computation Center (ASCC) and the allocation of supercomputer time at NERSC through grant number DE-AC02-05CH11231.  The work at LANL was supported by the U.S. DOE NNSA under Cont. No. 89233218CNA000001 through the LANL LDRD Program and the CINT, a DOE BES user facility.  We thank Adrian Feiguin for stimulating discussions. 

\section*{Author contributions}
R.S.M., B.S., C.L., and A.B. all contributed to the research reported in this study and the writing of the manuscript.

\section*{Additional information}
The authors declare no competing financial interests. 

\end{document}


\title{
--- Supplementary Information ---\\*[1em]
High-order Van Hove singularities in cuprates and related high-$T_c$ superconductors }

\author{Robert S. Markiewicz}
\email{r.markiewicz@northeastern.edu}
\affiliation{Department of Physics, Northeastern University, Boston, Massachusetts 02115, USA}

\author{Bahadur Singh}
\email{bahadur.singh@tifr.res.in}
\affiliation{Department of Condensed Matter Physics and Materials Science, Tata Institute of Fundamental Research, Colaba, Mumbai 400005, India}

\author{Christopher Lane}
\affiliation{Theoretical Division, Los Alamos National Laboratory, Los Alamos, New Mexico 87545, USA}
\affiliation{Center for Integrated Nanotechnologies, Los Alamos National Laboratory, Los Alamos, New Mexico 87545, USA}

\author{Arun Bansil}
\affiliation{Department of Physics, Northeastern University, Boston, Massachusetts 02115, USA}

\maketitle


\beginsupplement

\section{DOS calculations}
We now elaborate on the density of states (DOS) calculations. To determine the nature of the DOS divergence at the Van Hove singularity (VHS), the DOS calculations are carried out at $T=0$ to mimic the absence of disorder and nanoscale phase separation. Since the DOS is defined as a sum of delta-functions, we use a binning technique to minimize artificial broadening,
$$N(E) =\frac{2}{N_0}\sum_k\delta (E-\epsilon_k)\simeq\frac{2}{N_0}\sum_iN_i\delta(E-E_i),$$
where $N_0$ is the total number of $k$-points and the factor of 2 is used for the spin degeneracy. $N_i=\int_{E_i-\Delta/2}^{E_i+\Delta/2}N(E)dE$ is the number of $k$-states with energies $E_i=(i-1/2)\Delta$ in the $i^{th}$ bin.  Thus, as $\Delta\rightarrow 0$ and $N_0\rightarrow\infty$, $N_i(E=E_i)\rightarrow N(E)$.  We write $N_0 = (2N_k+1)^2$, where $N_k$ is the number of $k$-values along the positive $x-$ or $y-$axis. The DOS associated with dispersion of Eq. 1 of the main text is investigated for various values of $t-t'-t''$ with $0>t'/t>-0.5$, $0>t''/t'>-1$. Notably, the investigated values of $t-t'-t''$ correspond to exploring material space away from the cuprates (with $t''=-t'/2$) for hole doping. Reversing the sign of $t'$ and $t''$ will result in similar VHSs for electron-doping rather than hole-doping. 

\begin{figure}[h!]
\centering 
\includegraphics[width=0.5\textwidth]{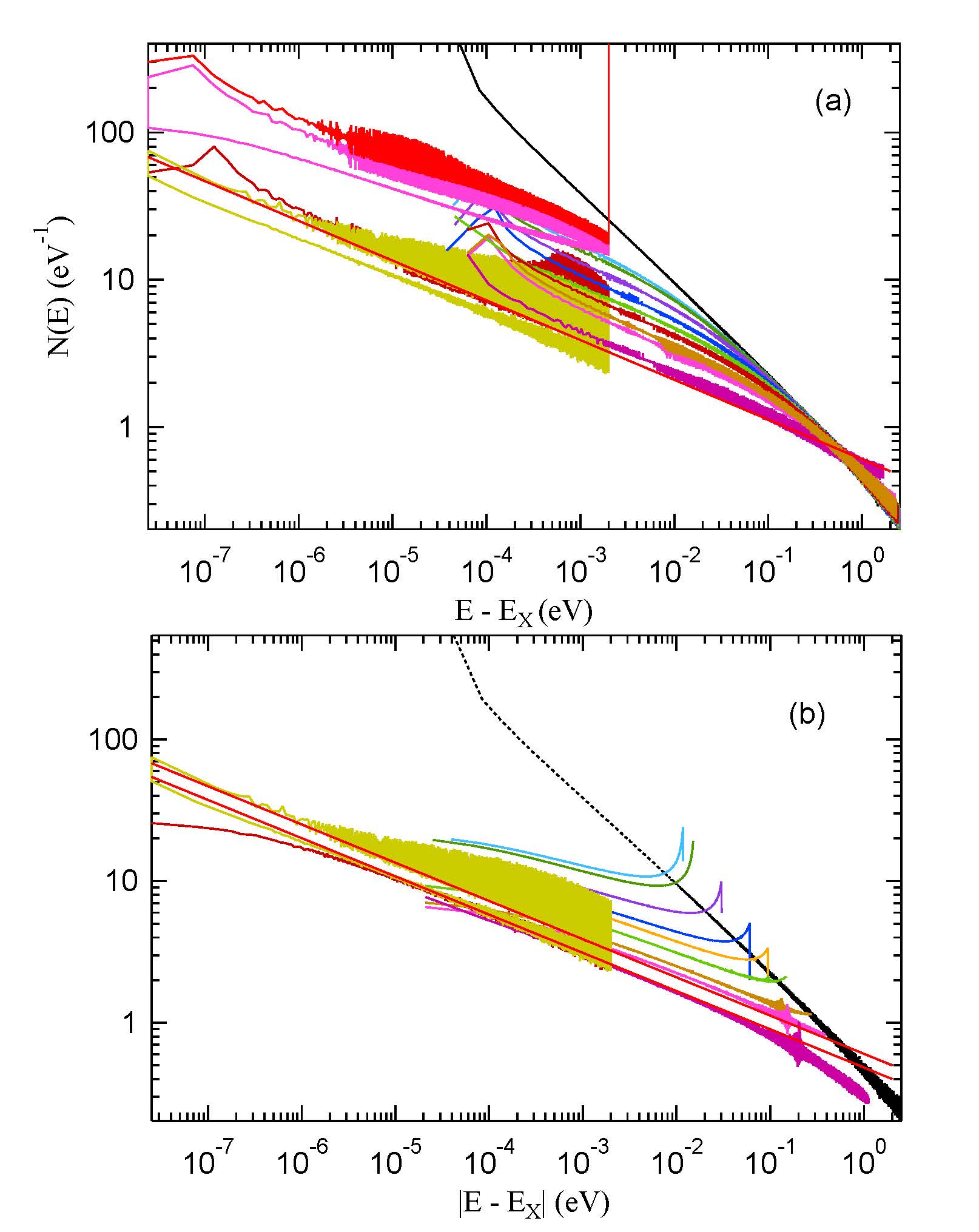}
\caption{{\bf Power-law divergence of $N(E)$ at $t'_c$.}  (a) N(E) at $t'_c$ for various $t''/t'$ values as shown in Fig. 2(a) of the main text but with an added higher-resolution data for $E-E_X < 2$ meV. (b) Same as (a) but for the singularity appearing at $ E-E_X < 0 $.}
\label{fig:A1c}  
\end{figure}

\begin{figure}[h!]
\centering 
\includegraphics[width=0.5\textwidth]{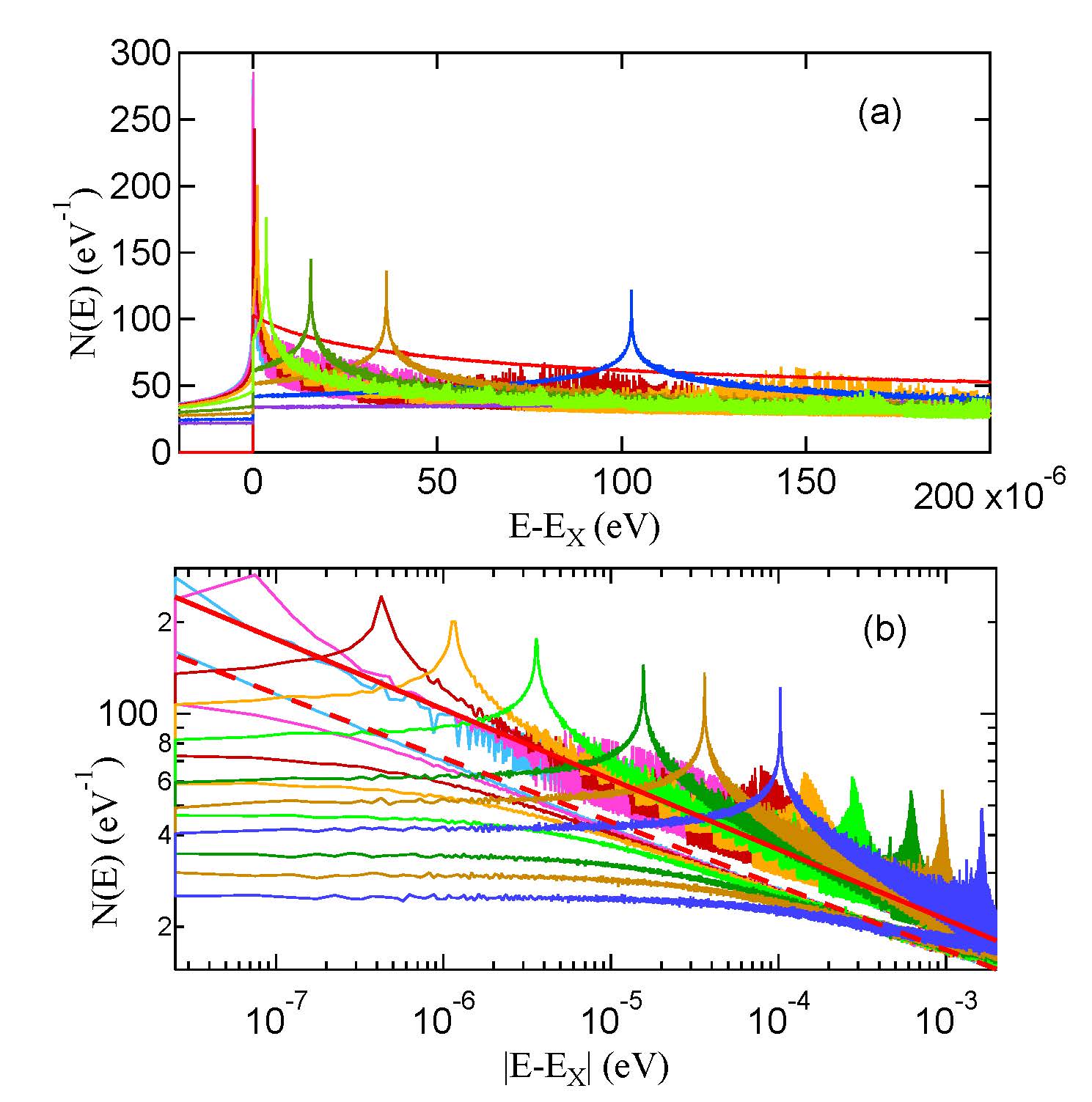}
\caption{{\bf Tuning $t'$ away from $t'_c$ for $t''/t'=-0.002$.}  Frame (b) is a log-log plot of the data of frame (a).  Note that in frame (b), each curve appears twice, the lower one for $E-E_X$ negative.  Both sets of curves display common initial power-law divergences before deviating at low $|E-E_X|$, with different slopes, $p$ = -0.23 (red solid curve) on positive side, -0.21 (red dashed curve) on negative side.}
\label{fig:A1b}
\end{figure}

We present DOS in \Fref{fig:A1c} at finer energy scale than in Fig.~2(a) using a smaller bin size with a denser $k$-mesh. The high-order VHSs are clearly resolved. It should be noted that the increased broadening of the curves is a sign of statistical error due to under $k$-point sampling. The problem is least in the range of small $\delta E$s, where $N(E)$ is largest and is unimportant for larger $\delta E$s, where larger bins can be used. In \Fref{fig:A1b} we show similar plots to Figs.~1(c,d) of the main text, but for different $t''/t'$ and finer bins, to illustrate the universality of the DOS variations.

\section{Van Hove dichotomy}

\subsection{$(\pi,\pi)$ VHS}

In the main text, we have considered VHS with diverging $q=0$ susceptibility. However, the VHS also has a diverging susceptibility at a second wave momentum $Q$ near $(\pi,\pi)$ as well. This is resolved in \Fref{fig:3} where the variation of $t'$ shifts the dominant susceptibility peak from $q=0$ to $q=(\pi,\pi)$. Similar effects arise with doping {\it i.e.} for $x \ne x_{VHS}$, each VHS evolves into a finite susceptibility peak with different doping dependences. This dichotomy is very relevant for cuprate physics since only the VHS near $(\pi,\pi)$ is correlated with the pseudogap crossover temperature, $T_Q\simeq T_{pg}$. We now demonstrate how these two VHSs compete and evolve with $t'$ and $x$.

\begin{figure}[h!]
\centering 
\includegraphics[width=0.5\textwidth]{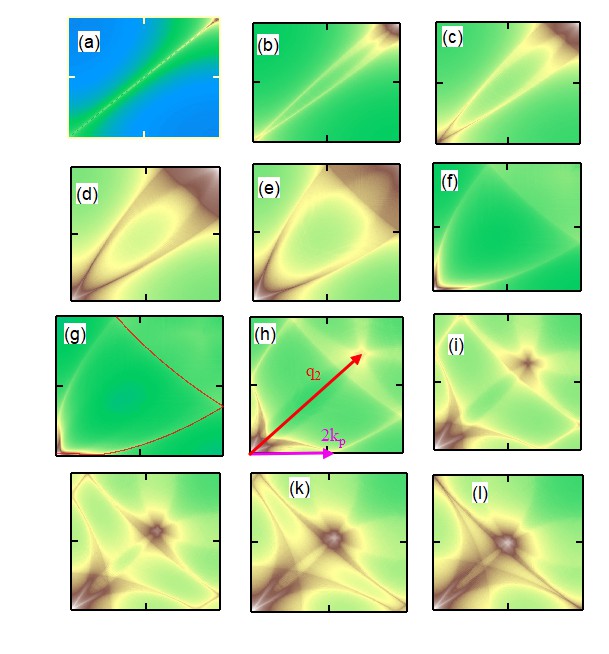}
\caption{Susceptibility maps at saddle-point VHS for $t'/t$ = (a) 0.0, (b) -0.05, (c) -0.10, (d) -0.15, (e) -0.20, (f) -0.25, (g) -0.258 [Bi2201], (h) -0.30, (i) -0.35, (j) -0.40, (k) -0.45, and (l) -0.50.   Blue and white identify minimum and maximum intensities, respectively.}
\label{fig:2} 
\end{figure}

\begin{figure}[h!]
\centering 
\includegraphics[width=0.5\textwidth]{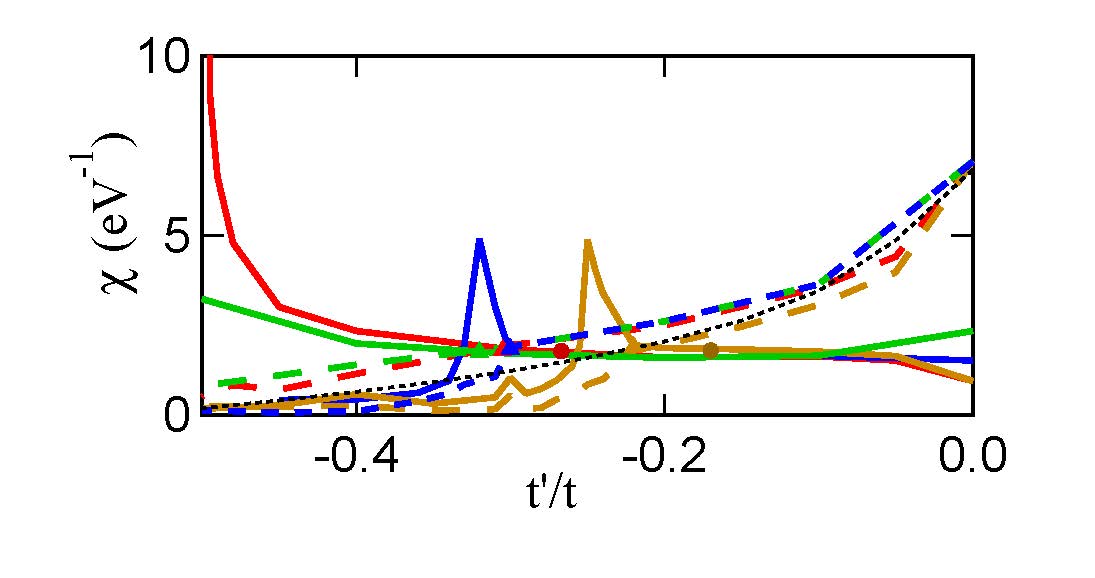}
\caption{ Competition between near-$(\pi,\pi)$ (dashed lines) and $\Gamma$-VHSs (solid lines), for four reference families, $t''/t'$ = 0 (red lines), -0.25 (blue lines), -0.5 (brown lines), or 1 (green lines).   Corresponding colored triangles indicate commensurate-incommensurate crossover.  For comparison, dark red ($t''=0$) and dark brown ($t''=-t'/2$) circles indicate the positions of the $x=0$ commensurate-incommensurate transitions discussed in Ref.~3 and Fig.~4 of the main text.}
\label{fig:3} 
\end{figure}

To capture the dichotomy, we calculate the full DFT-Lindhard susceptibility, 
\begin{equation}
\chi_0(q,\omega)=-\sum_k\frac{f(\epsilon_k)-f(\epsilon_{k+q})}{\omega -\epsilon_{k+q}+\epsilon_k}
\label{eq:1}
\end{equation}  The real part of $\chi_0(q,\omega =0)$ contains a folded ($q=2k_F$) map of the Fermi surface, whose intensity quantifies the strength of nesting at $2k_F$.  The susceptibility is not confined to the Fermi surface, but contains a substantial bulk contribution which can favor certain $k_F$ values, or even shift the peak to a near-by commensurate value, necessitating a careful numerical evaluation of $\chi_0$.$^3$  
   
Figure~\ref{fig:2} shows the evolution of the susceptibility map at the VHS doping as $t'$ is varied ($t''=-0.5t'$ is thought to be relevant to the cuprates). In \Fref{fig:2}(h), we show the Fermi surface nesting vector $q=2k_F$ (red line) over part of the first Brillouin zone. This clearly identifies that the nonanalytic features of the susceptibility are associated with the Fermi-surface nesting.  It can be seen that for each $|t'|$, there are susceptibility divergences at $\Gamma$ and  $(\pi,\pi)$.   However for small $|t'|$, the more intense VHS feature is at $(\pi,\pi)$ whereas for $t'< -0.18t$, the peak at $\Gamma$ becomes most intense.  For intermediate values of $|t'|$, there is an additional peak at $(\pi-\delta,\pi-\delta)$ (see \Fref{fig:2}(h) for example). Figure \ref{fig:3} shows intensities of the respective VHSs as a function of $t'$ for several values of $t''$.  


To understand the $(\pi,\pi)$ VHS evolution, we recall that for the pure Hubbard model ($t'=0$), the $(\pi,\pi)$-susceptibility has a $log^2T$ divergence that reduces to 
$$\chi((\pi,\pi),0)\sim log(T)log(T_X),$$ where $T_X=max\{T,T_{eh}\}$ and $k_BT_{eh}=|t'|$, for finite $t'$.\cite{VHS2} The dotted black line in \Fref{fig:3} is proportional to $ln(|t'|)$ (with an effective $t'$ at $t'=0$, to account for the finite $k$-mesh), showing that this is a good approximation over a broad range of $t'$, even in the presence of $t''\ne 0$.  For each pair of curves in \Fref{fig:3}, the crossover from $(\pi,\pi)$-dominated to $\Gamma$-dominated susceptibility is denoted by a triangle of the same color.  These tend to cluster near $t'=-0.3t$, but can be pinned by a nearby superVHS, as when $t''=-0.5t'$, brown triangle.  These crossovers closely scale with the commensurate-incommensurate [$(\pi,\pi)-(\pi,\pi-\delta)$] transition in the undoped cuprates, colored circles, which has been identified as the Mott-Slater transition.$^3$ 

We note from \Fref{fig:3} that judging by the DOS, $t'=0$ is the least likely place to look for strong instabilities, yet $\chi_0$ has its strongest divergence here.  Why are there two instabilities at each VHS?  The $q=\Gamma$ instability is associated with intra-VHS scattering, that at $(\pi,\pi)$ with inter-VHS scattering.  While the susceptibility at $\Gamma$ is constrained to be a Fermi surface effect, for general $q$ there can be a strong contribution to $\chi_0$ from states far from $E_F$, leading to peaks with very broad tails.  These tails play an important role near high symmetry points, when multiple Fermi surface sections approach the symmetry point.  Figure~\ref{fig:2} illustrates this effect for the $(\pi,\pi)$ plateau.   For larger $|t'|$, Figs. \ref{fig:2}(f)-(l), the weight is concentrated near the nesting map, $q=2k_F$.  However, as $|t'|$ decreases, overlap of the bulk spectral weight rapidly shifts the peak to exactly $(\pi,\pi)$, as the plateau shrinks down to a point for $t'=0$ (Figs. \ref{fig:2}(a)).  This has two effects- first, a prominent commensurate-incommensurate transition$^3$, and second, the rapid decrease of intensity at $(\pi,\pi)$ with increasing $|t'|$ seen in \Fref{fig:3}.  We note that the Fermi surfaces responsible for the $q\sim (\pi,\pi)$ plateau come from $k_F\sim (\pi/2,\pi/2)$ -- i.e., far from the VHS and hence not sensitive to $t''$.  Indeed, the $t'$-dependence of the $(\pi,\pi)$ plateau area, \Fref{fig:2}, is qualitatively similar to the doping dependence of the plateau area at fixed $t'$, \Fref{fig:1}.

Finally, the improved understanding of the $(\pi,\pi)$-VHS clears up a puzzle in Ref.~3.  There it was found (Fig. 6) that the susceptibility could be decomposed into two contributions, one associated with Fermi surface nesting, and a second, larger, contribution which is insensitive to the Fermi surface but peaked at $(\pi,\pi)$.  That peak can now be identified as due to VHS nesting\cite{RiSc}, with intra-VHS nesting near $\Gamma$ and inter-VHS nesting at, e.g.,  $Q=(\pi,0)-(0,-\pi)=(\pi,\pi)$.  In contrast, when the VHS is at the Fermi energy, there is also a Fermi-surface nesting contribution at $q=(\pi,0)-(-\pi,0)=(2\pi,0)$, which is equivalent to $\Gamma$.  Thus, VHS-nesting shifts the peaks in susceptibility closer to $(\pi,\pi)$ even when the Fermi level is away from the VHS peak.  

\section{Further topics}
In the remainder of this SM, we briefly discuss a few issues of relevance to the study of VHSs in real materials.

\subsection{Tuning away from the VHS}

\begin{figure}[h!]
\centering 
\includegraphics[width=0.5\textwidth]{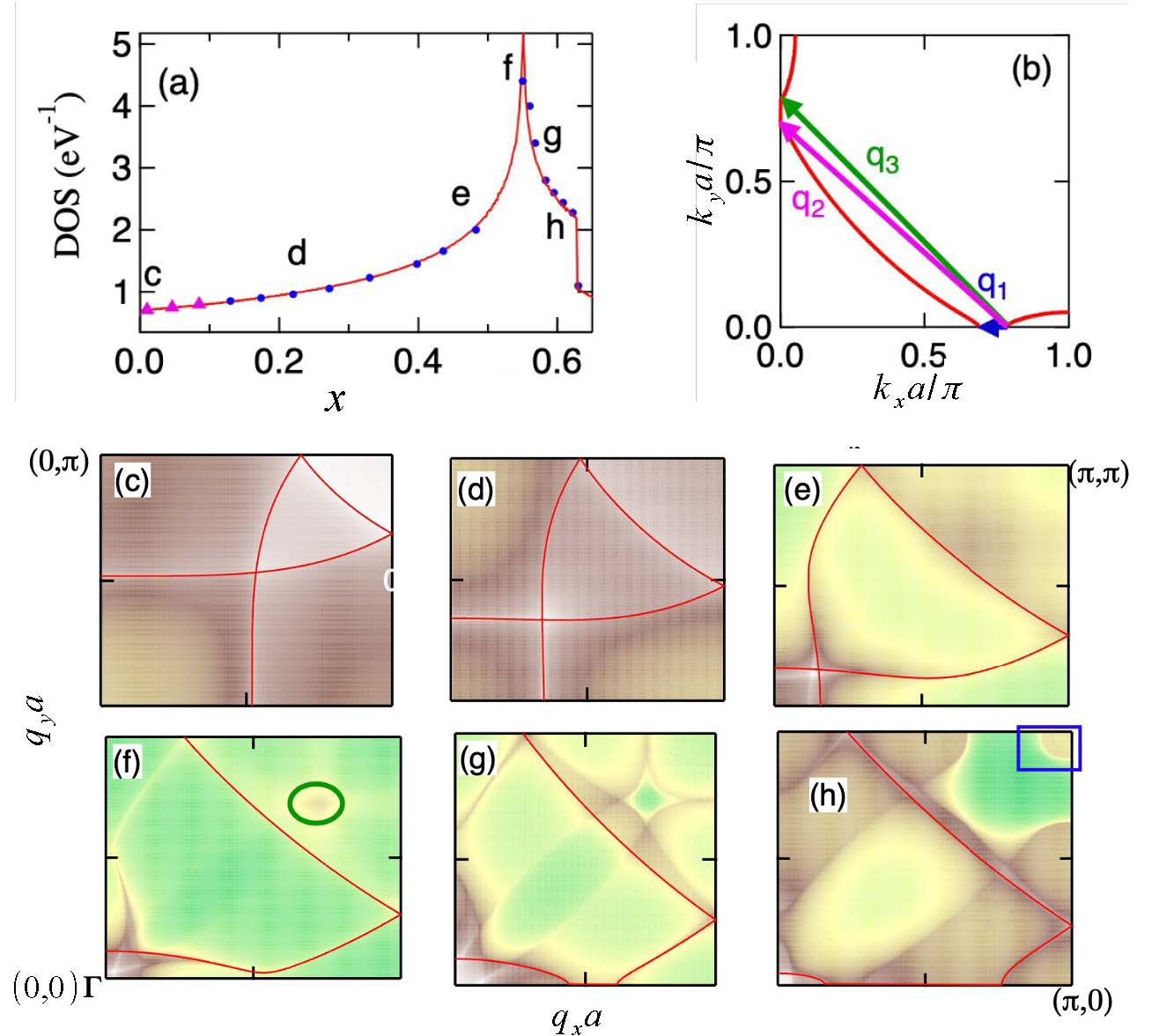}
\caption{{\bf Susceptibility maps showing evolution of nesting lines with doping for $t'/t=-0.3$, $t''/t'=-0.5.$} 
(a) DOS, showing doping of each susceptibility map. (b) Fermi surface for frame (g) showing three inter-surface nesting vectors.  (c-h) Susceptibility maps, with $q=2k_F$ nesting surface superposed as a red line; in frames (f-h), one branch is omitted to emphasize details of the nesting maps. Color scheme same as \Fref{fig:2}.  In frames (c-h), the $(\pi,\pi)$-plateau is defined as the area enclosed by the folded Fermi surface nearest to $(\pi,\pi).$}
\label{fig:1}
\end{figure}

While each dispersion leads to a unique $x_{VHS}$, the peaks in $\chi$ are controlled by the VHS over a wide doping range away from $x_{VHS}$.  For example, in Figure~\ref{fig:1} we illustrate the evolution of the bare susceptibility with doping for a typical $t'/t = -0.3$.  The susceptibility has a number of nonanalytic peaks, associated with Fermi surface nesting.  To illustrate this, we include superposed maps of the main barrel Fermi surfaces, doubled and folded back to the first Brillouin zone (BZ), to highlight the $q=2k_F$ nesting features.\cite{MLSB}  Additional features in frames (f-h) are associated with non-$2k_F$ nesting,\cite{MLSB,MGu}, e.g., inter-Fermi surface pocket nesting as illustrated in frame (b) for the doping of frame (g).  

To better understand the structure in these susceptibility maps, we briefly review the evolution of the Fermi surface, and its two topological transitions\cite{MLSB,MGu}.  Figure~\ref{fig:1}(a) shows the corresponding DOS, with points at which the susceptibility maps were sampled.  The DOS has two VHSs -- a logarithmic peak where a pocket pinches off near $(\pi,0)$, frame (f), and a step down when the pocket disappears with further hole doping, frame (h).  
Due to the folding, the pockets appear near $\Gamma$ in the susceptibility maps.  

For present purposes, our main result is to see how the VHS competition plays out as the Fermi level is shifted away from $E_{VHS}$.  As doping $x$ increases, the dominant nesting vector shifts away from $(\pi,\pi-\delta)$ at low doping (violet triangles in frame (a)) to $(\delta',\delta')$, or antinodal nesting (ANN), at higher doping (blue dots), near $x_{cross}\sim 0.12$.  As the VHS peak is approached, the ANN peak flows to $\Gamma$, $\delta'\rightarrow 0$ at the VHS, while the peak at $(\pi,\pi-\delta)$ evolves to a weak peak at $(\pi-\delta'',\pi-\delta'')$ at the VHS, green circle in \Fref{fig:1}(f).  Experimentally, the peak at $(\pi,\pi-\delta)$ is associated with an incommensurate SDW, while the ANN peak drives a CDW instability.  Thus we have demonstrated that this competition is controlled by the competition between intra-VHS scattering near $\Gamma$ and the inter-VHS scattering near $(\pi,\pi)$.$^{16}$  That is, the VHSs are important not only at $x_{VHS}$, but influence the dominant susceptibility over a wide doping range.  

The peak crossover seen at $x_{cross}=0.12$ in \Fref{fig:1} is reflected in a similar crossover of the VHS peaks, \Fref{fig:3}, at $t'_{cross}\sim -0.166$, for $t''=-0.5t'$.  For all $t'>t'_{cross}$ there is no $x_{cross}$ and the $(\pi,\pi)$-VHS is always dominant, while for $t'<t'_{cross}$ there is always an $x_{cross}$ where the influence of the $(\pi,\pi)$-VHS is lost.  Note that in the example of \Fref{fig:1}, $x_{cross}=0.12 << x_{VHS}=0.54$.

\subsection{New $ln^2T$ VHSs}
In this Subsection, we ask the whether the original Hubbard model, $t'=t''=0$, is unique in having a $ln^2(T)$ susceptibility divergence.  We demonstrate that it is not unique by generating a family of dispersions with this property.  As a byproducy, we demonstrate an improved technique for generating tight-binding models, in which the hopping parameters are approximately orthogonal.

\begin{figure}[h!]
\centering 
\includegraphics[width=0.5\textwidth]{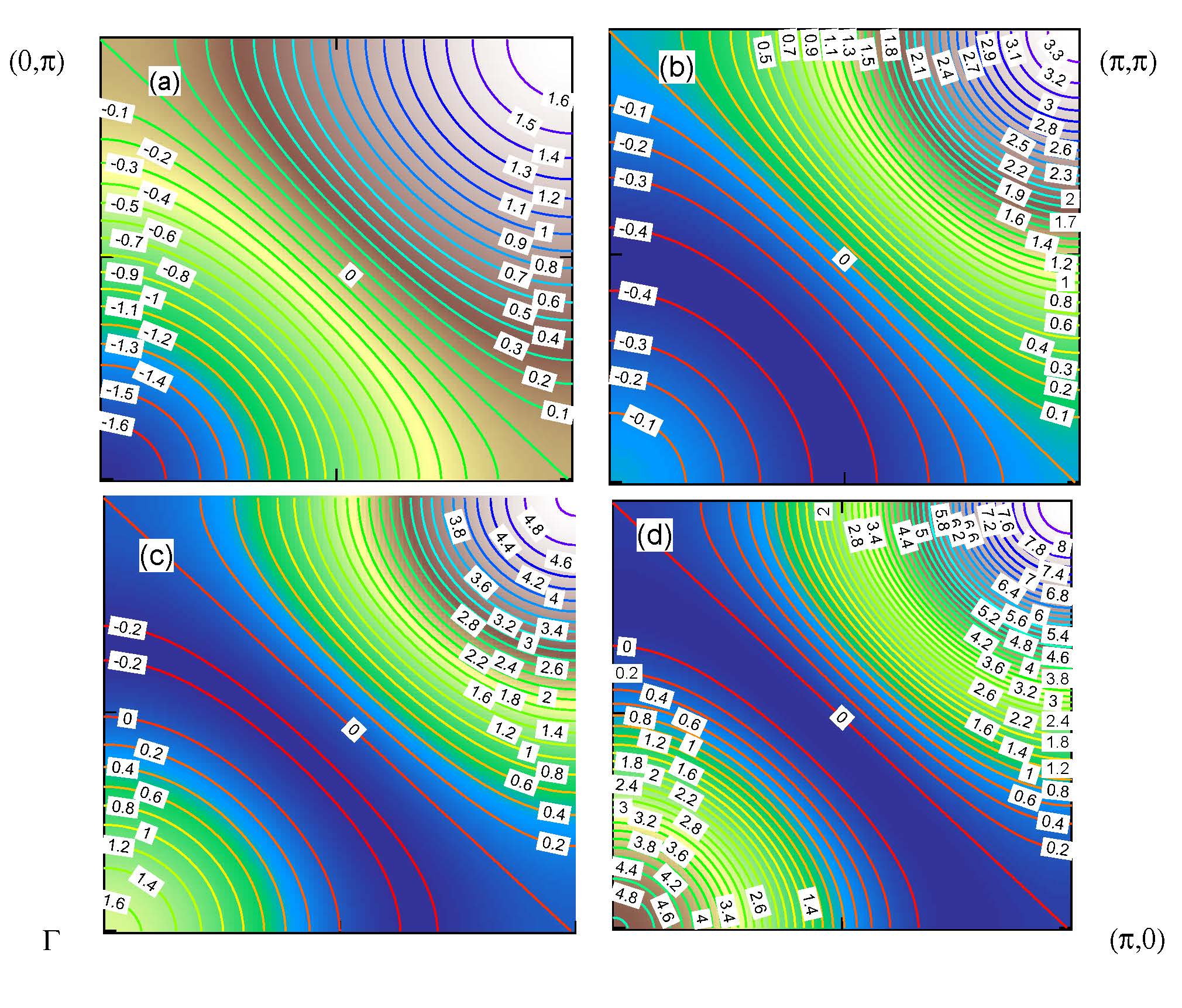}
\caption{{\bf Dispersion maps $E-E_X$ showing evolution of constant energy surfaces with $t'$ for $t''/t'=0.5.$}  $t'/t = $ (a) 0, (b) -0.5, (c) -1, and (d) -0.2.  Constant energy contours are labeled in eV.}
\label{fig:A8}
\end{figure}

Here, we generalize the analysis of Ref.~\onlinecite{VHS2} for finite $t''$.  The susceptibility divergence arises from the energy integral in Eq.~\ref{eq:1}, with $\epsilon_k$ near, e.g., $(\pi,0)$, while $\epsilon_{k+Q}$ is near $(0,\pi)$.  For the $t-t'-t''$ model, the $t'$ and $t''$ terms cancel, and the denominator becomes $\epsilon_k-\epsilon_{k+Q}=-4t(c_x+c_y)$.  Near $(0,\pi)$, $c_x+c_y\sim k_x^2-k_y^{'2} \sim k^2cos(2\phi)$, with $k_y'=\pi-k_y$.  Then the integral $\int kdk/k^2$ produces a logarithmic divergence, which can be cut off by, e.g., the temperature $T$. The angle integral $\int d\phi/cos(2\phi)$ also can diverge, if $\phi\rightarrow\pi/4$, leading to a second $ln(T)$ factor.   However, the fermi functions in Eq.~\ref{eq:1} can cut off this divergence, since at low $T$ they only allow scattering from full to empty states, and hence cut the $\phi$ integral off at the Fermi surface.  To test this possibility, we rewrite Eq.~1 by using $cos(2\theta)=2cos^2(\theta)-1$, as
\begin{equation}
\epsilon_k=-2t(c_x+c_y)-(t'+2t'')(c_x+c_y)^2+(t'-2t'')(c_x-c_y)^2$$\vskip 0.1in$$
+4t''.
\label{eq:4}
\end{equation}
Thus, when $t''\ne 0$, the cutoff of Eq.~\ref{eq:4} becomes $k_BT_{eh}=|t'-2t''|$, and when $2t''=t'$, the susceptibility has a $ln^2(T)$-divergence independent of $t'+t''$, leading to a new reference family of anomalous VHSs, \Fref{fig:A8}.

Figure~\ref{fig:A8} shows that the dispersion changes strongly with $t'$, whereas the contours of constant energy are independent of $t'$, since they represent $c_x+c_y=R=constant$.  Consequently, for all values of $t'$, the zone diagonal $(0,\pi)\rightarrow (\pi,0)$ lies at the VHS, $E_F=E_X$, ensuring a square Fermi surface and a $ln^2(T)$-divergence of the $(\pi,\pi)$ susceptibility.  However, there is still a competition of $(\pi,\pi)$ and $\Gamma$ susceptibilities, except now in terms of high-order VHSs.  Note that for $t'=0$ (the conventional Hubbard model) the dispersion is electron-hole symmetric.  For $t'<0$, the dispersion becomes asymmetric, with a local peak developing at $\Gamma$, leading to a Mexican hat potential, which moves toward the $(0,\pi)\rightarrow (\pi,0)$ line as $|t'|$ increases.  The minimum of the Mexican hat produces a flat band quite similar to that found in secondary high-order VHSs.  

This similarity is not accidental.  For a mean-field model of $Q=(\pi,\pi)$ AFM order, the dispersion in the AFM phase becomes $E_{\pm}= \epsilon_+\pm\sqrt{\epsilon_-^2+U^2/4}$, where $\epsilon_{\pm}=(\epsilon_k\pm\epsilon_{k+Q})/2$.  When $U$ is large, the square-root term becomes $U/2 +\epsilon_-^2/U$.  In the $t-t'-t''$ families, $\epsilon_-=-2t(c_x+c_y)$, and the square-root term becomes $U/2 +J(c_x+c_y)^2$, $J=4t^2/ U$, having the same nonlinear term as Eq.~\ref{eq:4}.

Thus, remarkably, whereas the DOS is highly sensitive to the Fermi surface, the $q=(\pi,\pi)$ susceptibility obeys the simpler result of Eq.~\ref{eq:4}, and can only have $ln$ or $ln^2$ divergences in the $(\pi,\pi)$ susceptibility.  We note that the original Hubbard model required extreme fine tuning (all hopping parameters except $t$ equal to zero), calling into question the relevance of a square Fermi surface and resultant $ln^2$ divergence.  By showing that a square Fermi surface can be found along an entire cut in parameter space, we have greatly enhanced the likelihood that such a feature, with accompanying high-order VHSs, might be experimentally observed.  

Finally, Ref.~3 introduced the notion of reference families,  approximate dispersions with only a few hopping parameters, which approximately capture the same physics as more accurate dispersions with considerably more parameters.  Reference families are useful in genomic studies, seeing what materials share the same referene families and allowing one to 'tune' the parameters between different materials.  For these families to be of value, it is necessary that higher-order hopping parameters have negligible effects in most cases.  This new, `Hubbard' family offers a simple test: can a line of square-Fermi-surface states persist as higher order hopping parameters are included?  We explore this issue by looking at the most general TB model with terms of order $c_i^3$ ($E_3$) and $c_i^4$ ($E_4$), $\epsilon_{k,4} = \epsilon_k +E_3 +E_4$.  Here $\epsilon_k$ is given by Eq~1 and
\begin{equation}
E_3=-2t_3(c_{3x}+c_{3y})-4t_4(c_{2x}c_y+c_{2y}c_x)$$\vskip 0.1in$$
=2a_3(c_x+c_y)-4A_3(c_x+c_y)^3-4B_3(c_x-c_y)^3,
\label{eq:4a}
\end{equation}
with $a_3=t_3+6t_4$, $A_3,B_3=t_3\pm t_4/3$,
\begin{equation}E_4=-2t_5(c_{4x}+c_{4y})-4t_6(c_{3x}c_y+c_{3y}c_x)-4t_7c_{2x}c_{2y}
$$\vskip 0.1in$$
=e_{40}-a_4(c_x+c_y)^2-b_4(c_x-c_y)^2-2A_4(c_x+c_y)^4-2B_4(c_x-c_y)^4
$$\vskip 0.1in$$
-4C_4(c_xc_y)^2,
\label{eq:4b}
\end{equation}
with $e_{40}=-4(t_5+t_7)$, $a_4,b_4=4(t_5+t_7)\mp 3t_6$, $A_4,B_4=4t_5\pm t_6$, $C_4=4(t_7-6t_5)$.  Finally, since $c_xc_y=[(c_x+c_y)^2-(c_x-c_y)^2]/4$, $\epsilon_{k,4}$ can be written as a polynomial in $(c_x+c_y)$ and $(c_x-c_y)$.  Eliminating all the terms in $(c_x-c_y)$ leaves behind a polynomial in $(c_x+c_y)$, a 4th order generalization of the Hubbard reference family.

We note in passing that our new series expansion, a Taylor series in the two variables $X=c_x+c_y$ and $Y=c_x-c_y$, should be a significant improvement over the conventional neighbor-by-neighbor tight-binding model.  We see from Eqs.~\ref{eq:4a} and~\ref{eq:4b} that each additional neighbor introduces two kinds of correction: new terms in $X$ and $Y$ (terms whose coefficients are capital letters on right-hand side of equations) and renormalizations of lower-order terms (terms with lower-case coefficients).  In contrast, the new series is essentially a symmetry-corrected Fourier series, ensuring that the new terms added at each order are orthogonal to the lower-order terms, and will be accepted only if they genuinely improve the fit to the dispersion.  Moreover, the even order terms are electron-hole symmetric while odd-order terms are not.  Only the latter contribute to shifting the VHS from half-filling.